\documentclass[11pt,a4paper]{article}
\usepackage{jheppub,amsmath, amsthm, amssymb,slashed,url}
\usepackage{graphicx}
\usepackage{epstopdf}
\usepackage[all]{xy}
\usepackage{multirow}

\font\teneurm=eurm10 \font\seveneurm=eurm7 \font\fiveeurm=eurm5
\newfam\eurmfam
\textfont\eurmfam=\teneurm \scriptfont\eurmfam=\seveneurm
\scriptscriptfont\eurmfam=\fiveeurm

\font\teneusm=eusm10 \font\seveneusm=eusm7 \font\fiveeusm=eusm5
\newfam\eusmfam
\textfont\eusmfam=\teneusm \scriptfont\eusmfam=\seveneusm
\scriptscriptfont\eusmfam=\fiveeusm

\font\tencmmib=cmmib10 \skewchar\tencmmib='177
\font\sevencmmib=cmmib7 \skewchar\sevencmmib='177
\font\fivecmmib=cmmib5 \skewchar\fivecmmib='177
\newfam\cmmibfam
\textfont\cmmibfam=\tencmmib \scriptfont\cmmibfam=\sevencmmib
\scriptscriptfont\cmmibfam=\fivecmmib
\def\pa{\partial}

\newcommand{\mc}[1]{\mathcal{#1}}

\newcommand{\si}{\sigma}
\newcommand{\lam}{\lambda}
\newcommand{\olam}{\overline{\lambda}}
\newcommand{\g}{\gamma}
\newcommand{\e}{\epsilon}

\newcommand{\ophi}{\overline{\phi}}
\newcommand{\opsi}{\overline{\psi}}
\newcommand{\oF}{\overline{F}}
\newcommand{\oep}{\overline{\e}}

\newcommand{\Sp}{\mathbb{S}}
\newcommand{\Rp}{\mathbb{RP}}
\newcommand{\compe}{\vartheta}
\newcommand{\compt}{\varphi}
\newcommand{\compd}{y}

\newcommand{\ch}{\text{\boldmath $Q$}}

\renewcommand{\include}[1]{}
\renewcommand\documentclass[2][]{}

\title{Varieties of Abelian mirror symmetry on $\mathbb{RP}^2 \times \mathbb{S}^1$}

\preprint{\begin{flushright} OU-HET 884 \\ RIKEN-STAMP-21 \end{flushright}}
\author[a]{Hironori Mori}
\author[b]{and Akinori Tanaka}
\affiliation[a]{Department of Physics, Graduate School of Science, Osaka University, Toyonaka, Osaka \\560-0043, Japan}
\affiliation[b]{iTHES Research Group, RIKEN, Wako, Saitama \\351-0198, Japan}
\emailAdd{hiromori@het.phys.sci.osaka-u.ac.jp}
\emailAdd{akinori.tanaka@riken.jp}

\abstract{
We study 3d mirror symmetry with loop operators, Wilson loop and Vortex loop, and multi-flavor mirror symmetry through utilizing the $\Rp^2 \times \Sp^1$ index formula.
The key identity which makes the above description work well is the mod 2 version of the Fourier analysis, and we study such structure, the S-operation in the context of a SL$(2,\mathbb{Z})$ action on 3d SCFTs. We observed that two types of the parity conditions basically associated with gauge symmetries which we call $\mathcal{P}$-type and $\mathcal{CP}$-type are interchanged under mirror symmetry. We will also comment on the T-operation.
}
\begin{document} \maketitle

%!TEX encoding = JISJapanese

\include{common_supp_files/sonota/begin2}
%%%%%%%%%%%%%%%%%%
%%%%%%%%%%%%%%%%%%
\section{Introduction}
Three-dimensional $\mathcal{N} = 2$ mirror symmetry \cite{Intriligator:1996ex, Aharony:1997bx, Kapustin:1999ha, Tong:2000ky} is one of the nontrivial classes of dualities among quantum field theories, which states that two distinct  theories defined in the UV regime flow into an identical IR fixed point.
In order to test such dualities, exact physical quantities protected by supersymmetry are often useful and provide to us understandings of the duality in mathematically well-defined languages.

In this paper, we deepen the study of 3d Abelian mirror symmetry on $\mathbb{RP}^2 \times \mathbb{S}^1$ performed in \cite{Tanaka:2014oda, Tanaka:2015pwa}.
We show that the correspondences of loop operators and the multi-flavor version of mirror symmetry which were studied on other three-dimensional manifolds \cite{Kapustin:2011jm, Kapustin:2012iw, Drukker:2012sr} also work even on $\mathbb{RP}^2 \times \mathbb{S}^1$ in terms of the index \cite{Imamura:2011su}. %which can be exactly computed by the localization \cite{Pestun:2007rz}.
The key mathematical identity to demonstrate them is a \textit{mod 2} delta function's formula:
\begin{align}
%\frac{1}{2} (1 + 1) =1,
%\quad 
%\frac{1}{2} (1 - 1) =0 
%\qquad
%\Leftrightarrow
%\qquad
\frac{1}{2} 
\sum_{a =0,1} 
e^{i \pi a (A-B) }  = \delta_{A,B}^{\text{mod 2}}.
\label{key}
\end{align}
This is nonvanishing only if $A - B \equiv 0$ (mod $2$). See \eqref{mod2Kdelta} for details.
The mod 2 structure can be regarded as an algebraic representation of the $\mathbb{RP}^2$ unorientable structure.

As considered somewhat systematically in \cite{Tanaka:2015pwa}, in order to define supersymmetric gauge theories on $\mathbb{RP}^2 \times \Sp^1$, there are \textit{two} types of parity conditions for the vector multiplet consistent with the unorientable structure which we call $\mc{P}$-type and $\mc{CP}$-type. The covariant derivative under each parity condition transforms as %and we represent it by using nodes with these characters in the quiver diagram:
\begin{align}
& \bullet \text{ $\mc{P}$-type vector multiplet $\ni A_{\mu}$ : } 
\quad
%\xymatrix@C=6pt@R=15pt {
%*++[o][F-]{\text{\scriptsize$\mc{P}$}} 
%}
%\quad \
(\pa_\mu - i \ch A_\mu) \to \pm(\pa_\mu - i \ch A_\mu).
\\
& \bullet \text{ $\mc{CP}$-type vector multiplet $\ni A_{\mu}$ : }
%\xymatrix@C=6pt@R=15pt {
%*++[o][F-]{\text{\tiny$\mc{CP}$}} 
%}
\ 
(\pa_\mu - i \ch A_\mu) \to \pm(\pa_\mu + i \ch A_\mu).
\end{align}
%Multi-flavored mirror symmetry can be represented by the following quiver diagrams:
%
%%%%%%%
%We show these correspondences including parity types by utilizing $\mc{N}=4,N_f=1$ mirror symmetry \cite{Intriligator:1996ex, Kapustin:1999ha} and the localization computations \cite{Tanaka:2014oda, Tanaka:2015pwa}.
%This is a counter part of the ones on $\Sp^2 \times \Sp^1$ observed by \cite{Krattenthaler:2011da, Kapustin:2011jm}.
The former simply brings the parity ($\mathcal{P}$) effect, while the latter accompanies charge conjugation ($\mathcal{C}$), $+ \ch \leftrightarrow - \ch$, with parity.

In addition, we observe that it is possible to calculate $e$-charged \textit{BPS-Wilson loop} $W_e$ and $m$-vorticity \textit{BPS-Vortex loop} $V_m$.
As well known, the Wilson loop is an order parameter of the system.
On the other hand, the Vortex loop, which is a codimension-2 defect in 3d theories, is a disorder parameter of the system and is defined by not a standard operator insertion but singular behavior of the fields around the loop \cite{Kapustin:2012iw, Drukker:2012sr, Assel:2015oxa}. In fact, there is one simple way \cite{Kapustin:2012iw} to describe the Vortex loop by making use of \textit{S-operation} of a SL$( 2, \mathbb{Z} )$ group in 3d CFTs \cite{Witten:2003ya}.
%When it works, we can write down the Vortex loop with vorticity $m$ as
%\begin{align}
%V_m=S^{-1} W_m S
%\end{align}
%In this case, the S-operation makes the field as singular around the $m$-charged Wilson loop after integrating out the additional fields which turned on when the S-operation performed.
We systematically find the correspondences of these loop operators based on that fact in the context of 3d $\mathcal{N} = 2$ mirror symmetry with a single matter $N_{f} = 1$:
\begin{align}
 \mc{P}\text{-type Gauge} 
 \left\{ \begin{array}{ll}
\text{Wilson}  \\
\text{Vortex}\\
\end{array} \right\}
 \text{loop} 
 \quad &\Leftrightarrow \quad 
 \mc{CP}\text{-type Global}
 \left\{ \begin{array}{ll}
 \text{Vortex} \\
 \text{Wilson} \\
\end{array} \right\}
 \text{loop}
\label{mirrorloopP1}
\\
 \mc{CP}\text{-type Gauge}
  \left\{ \begin{array}{ll}
\text{Wilson}  \\
\text{Vortex}\\
\end{array} \right\}
 \text{loop} 
\quad  &\Leftrightarrow \quad
\mc{P}\text{-type Global}
 \left\{ \begin{array}{ll}
\text{Vortex}  \\
\text{Wilson}\\
\end{array} \right\}
 \text{loop}
\label{mirrorloopCP1}
\end{align}
The Wilson loop on one side is mapped to the Vortex loop on the dual side as seen in the general discussion \cite{Kapustin:2012iw, Drukker:2012sr} and also in \cite{Assel:2015oxa} for $\mathcal{N} = 4$. The remarkable point in this duality is to interchange $\mc{P}$-type and $\mc{CP}$-type for the dynamical (Gauge) or background (Global) gauge field. This can be regarded as the non-trivial effect originated from the unorientable structure.

We also find that the mirror symmetry exchanges $\mc{P}$ and $\mc{CP}$ in many ways.
One example is multi-flavor mirror symmetry, the duality between the SQED with $N_f$ flavors (SQED$_{N_f}$) and U$(1)^{N_f -1}$ quiver gauge theory, which is realized by the following two ways:
\begin{align} \label{mflavor1}
	\begin{aligned}
	%\bullet 
        \text{ $\mc{P}$-SQED$_{N_f}$}
        \quad
        %&&
        & \Leftrightarrow
        \quad
        \text{ $\mc{CP}$-U$(1)^{N_f-1}$ gauge theory},
        \\
        %\bullet 
        \text{ $\mc{CP}$-SQED$_{N_f}$}
        \quad
        %&&
        & \Leftrightarrow
        \quad 
        \text{ $\mc{P}$-U$(1)^{N_f-1}$ gauge theory},
	\end{aligned}
\end{align}
where leading letters represent the parity type for the vector multiplet in the gauge group. These are new properties in 3d theories equipped with geometrical $\mathbb{Z}_2$-structure.

%\vspace{.5cm}
The rest of this paper is organized as follows.
In Section \ref{Localization}, we briefly review the exact results by localization techniques applied to path integrals of Abelian SUSY gauge theories on $\Rp^2 \times \Sp^1$ and provide recipe for the computation and the usage of S-operation.
In Section \ref{Mirrorloops}, we show that the correspondences of the BPS loop operators \eqref{mirrorloopP1} and \eqref{mirrorloopCP1} in single-flavor mirror symmetry by utilizing S-operation on $\Rp^2 \times \Sp^1$.
In Section \ref{Multiflavor}, we check multi-flavor mirror symmetry \eqref{mflavor1} by using the localization recipe and the gauging procedure of the global symmetry.
We comment on open questions in Section \ref{Conclusion}.
In Appendix, we summarize the bases of the BPS loop operators which are $\mathbb{Z}_2$-invariant on $\Rp^2$.

%%%%%%%%%%%%%%%%%%
%%%%%%%%%%%%%%%%%%
\include{common_supp_files/sonota/sections_end}

%!TEX encoding = JISJapanese

\include{common_supp_files/sonota/begin2}
%%%%%%%%%%%%%%%%%%
%%%%%%%%%%%%%%%%%%
\section{Features of SUSY gauge theory on $\mathbb{RP}^2 \times \mathbb{S}^1$} \label{Localization}
\subsection{Lightning review on localization computation on $\mathbb{RP}^2 \times \mathbb{S}^1$}
%%%

%\subsection{S-operation on $\mathbb{RP}^2 \times \mathbb{S}^1$}
This subsection is dedicated to the review of \cite{Tanaka:2014oda, Tanaka:2015pwa}. 
What we consider is $\mathbb{RP}^2 $ {index},
\begin{align}
\mc{I}^{\mathbb{RP}^2} (q)
=
\text{Tr}_{\mc{H}^{\mathbb{RP}^2}}
\Big[
(-1)^{\hat{F}}
q^{\frac{1}{2} (\hat{H} + \hat{j}_3) }
%\alpha^{\hat{f}}
\Big],
\label{qinddef}
\end{align}
where $q$ is fugacity and
$\hat{F}$, $\hat{H}$, $\hat{j}_3$ %, and $\hat{f}$ 
are fermion number, Hamiltonian, 3rd angular momentum. %and charge of a global symmetry.
$\mc{H}^{\mathbb{RP}^2}$ is the theory's $\mathbb{Z}_2$-divided Hilbert space.
In order to compute this quantity, we rewrite this definition by using path integral over the supersymmetric  multiplets, 
%Once we write it in path integral formalism, we can utilize an exact method, the celebrated localization technique 
%We can also introduce gauge field into consideration in supersymmetric way.
%There are two possibilities for supersymmetric multiplet in our situation:
\begin{align}
%\left. \begin{array}{ll}
%\text{: constructed by spin 0 $(\phi, \ophi)$ \& spin 1/2 $(\psi,\opsi)$ fields,}
\text{Vector multiplet} \ V=(\si, \lam, \olam, A_\mu),
\quad
\text{Matter multiplet} \ \Phi=(\phi, \psi), (\ophi, \opsi), 
%\text{: constructed by spin 0 $(\si)$, 1 $(A_\mu)$ \& spin 1/2 $(\lam, \olam)$ fields,}
% \\
%\end{array} \right.
\label{mults}
\end{align}
where $\phi$ and $\ophi$ are complex scalars, $\si$ is real scalar, $\psi$ and $\lam$ are spinors, $A_\mu$ is gauge field\footnote{
Vector multiplet $V$ include gauge field $A_\mu$, so the theory with $V$ becomes gauge theory automatically. Throughout this paper, we only consider U(1) gauge symmetry.
} .
The effect of considering $\mathbb{Z}_2$-divided Hilbert space should emerge in the boundary conditions for the fields under the reflection, and the \eqref{qinddef} becomes
\begin{align}
\mc{I}^{\mathbb{RP}^2} (q)
=
\int_{\mathbb{RP}^2 \times \mathbb{S}^1}
 \mc{D} \Phi 
 \mc{D}V
 \ e^{-S[\Phi,V]}
 ,
 \label{pi}
\end{align}
with appropriate boundary conditions as explained below.
In order to represent arbitrary theory's $\mc{I}^{\mathbb{RP}^2} (q)$, we use quiver diagram.
\paragraph{Vector multiplet}
Compared with the conventional notation, $\mathbb{RP}^2 \times \mathbb{S}^1$ quiver diagram is slightly different.
One important difference is that we have \textit{two types} of the vector multiplets $V^{(\mc{P})}$ and $V^{(\mc{CP})}$ \cite{Tanaka:2015pwa}, and represent them as nodes with characters $\mc{P}$ or $\mc{CP}$:
\begin{align}
\left. \begin{array}{lllll}
V^{(\mc{P})}:
&
\text{dynamical} &
=\xymatrix {*++[o][F-]{\text{\scriptsize$\mc{P}$}} },& 
\
\text{background}&
=\xymatrix {
*++[o][F.]{\text{\scriptsize$s^\pm, \theta$}}
}
\text{ where }
\Big(
\left. \begin{array}{ll}
s^+=0,  \\
s^- = 1,  
\end{array} \right.
\
\theta \in %\mathbb{S}^1_{2\pi}
[0, 2\pi]
\Big).
\\ %%%%%%
V^{(\mc{CP})}:
&
\text{dynamical} & 
=\xymatrix {*++[o][F-]{\text{\tiny$\mc{CP}$}} },&
\
\text{background} &
=\xymatrix {
*++[o][F.]{\text{\scriptsize$s,\theta_\pm$}} 
}
\text{ where }
\Big(
s \in  \mathbb{Z},
\quad
\left. \begin{array}{ll}
\theta_+=0  \\
\theta_- =\pi
\end{array} \right.
\Big).
\\
\end{array} \right.
%\bullet 
\label{vdof}
\end{align}
%where $A_{\text{flat}}$ is the nontrivial flat connection on $\Rp^2$, $A_{\text{mon}}$ is the Dirac monopole configuration with monopole charge $2$, and $\theta$ is a AB phase along $\mathbb{S}^{1}$.
$s_\pm$ represents holonomy along nontrivial path in $\pi_1(\mathbb{RP}^2)$, and $s$ corresponds to Dirac monopole charge\footnote{
$s_\pm = B_\pm$ and $s= B/2$ in \cite{Tanaka:2015pwa} notation.
} on $\mathbb{RP}^2$.
$\theta$ and $\theta_\pm$ are holonomy along $\mathbb{S}^1$.
%%%%%%%%%%%%%%%%%%%%%%%%%%%%
\paragraph{Matter multiplet}
We represent matter multiplet degrees of freedom by a rectangular node.
As same as the case of vector multiplet, we have basically two possible matter multiplets, $\Phi_s$ and $\Phi_d$, a singlet and a doublet:
\begin{align}
\left. \begin{array}{ll}
\Phi_s : &
\text{dynamical}
=
 \xymatrix {*+[F-]{ \Phi }}\ .
\\
\
\\
\Phi_d : &
\text{dynamical}
=
\xymatrix{
*+[F-]{^{\Phi_1}_{\Phi_2}}
}
\ .
\\
\end{array} \right.
\end{align}
For more detail, see \cite{Tanaka:2015pwa}.

%%%%%%%%%%%%%%%%%%%%%%%%%%%%
\newpage
\paragraph{Gauge coupling}
We represent gauge coupling between matter and vector multiplets with arrow $\to$ connecting the two nodes.
The number of arrowheads means charge $\ch$, and the direction of the arrow means the sign$(\ch)$.
To apply \eqref{qinddef}, we should make theory $\mathbb{Z}_2$-invariant at least classically, and the possible couplings are as follows:
\begin{align}
&%\bullet 
\xymatrix{
*+[F-]{\Phi}
\ar@{->}[r]^<{}_{\underbrace{}_{\ch}} 
&
*++[o][F-]{\text{\scriptsize$\mc{P}$}} 
},
\quad
\xymatrix{
*+[F-]{^{\Phi_1}_{\Phi_2}}
\ar @<1mm> [r]
\ar @<-1mm> [r]_{\underbrace{}_{\ch}}
&
*++[o][F-]{\text{\scriptsize$\mc{P}$}} 
},
\quad
\xymatrix{
*+[F-]{^{\Phi_1}_{\Phi_2}}
\ar @<1mm> [r]
&
*++[o][F-]{\text{\tiny$\mc{CP}$}} 
\ar @<1mm> [l]^{\underbrace{}_{\ch}}
}.
\end{align}
Note that matter singlet cannot couple to $\mc{CP}$-type because it includes charge conjugation $\mathcal{C}$, and singlet is not invariant under the $\mathcal{C}$. 
%It comes from charge conjugation.
Of course, one can couple the background vector multiplets in \eqref{vdof} in these way, and in such case, we will use dotted arrow $
\xymatrix{
\ar@{.>}[r]
&
}
$. 
%%%%%%%%%%%%%%%%%%%%%%%%%%%%
\paragraph{Topological coupling}
We can define $\mathbb{Z}_2$-invariant topological coupling only between $\mc{P}$-type vector multiplet and $\mc{CP}$-type vector multiplet,
\begin{align}
&
\xymatrix{
*++[o][F-]{\text{\scriptsize$\mc{P}$}} 
\ar@{.}[r]^{\text{BF}}
&
*++[o][F-]{\text{\tiny$\mc{CP}$}} 
},
\label{Topqui}
\end{align}
where we use the index ``BF" as positive coupling.
If the coupling is negative, we use the index ``$-$BF".
This coupling will be important in later discussion.
One can consider one of them as background factor multiplet, too.

%%%%%%%%%%%%%%%%%%%%%%%%%%%%
\paragraph{Superpotential}
We can even turn on a supersymmetric potential among matter multiplets called superpotential if the interaction does not break the $\mathbb{Z}_2$-invariance.
However, the final result does not depend on the form of superpotential, so we neglect them.

%\vspace{1cm}\noindent
For example, if we have the following quiver diagram, 
\begin{align}
&%\bullet 
\xymatrix{
*+[F-]{\Phi}
\ar@{->}[r]^<{}_{\underbrace{}_{\ch}} 
&
*++[o][F-]{\text{\scriptsize$\mc{P}$}} 
\ar@{.}[r]^{\text{BF}}
&
*++[o][F.]{\text{\scriptsize$s,\theta_\pm$}} 
}
\label{example1}
\end{align}
then, what we calculate is the following quantity:
\begin{align}
\mc{I}^{\mathbb{RP}^2}
(q; \ s, \theta_\pm)
=
\int_{\mathbb{RP}^2 \times \mathbb{S}^1}
\mc{D} \Phi
\mc{D} V^{(\mc{P})}
\
e^{-S},
\label{example2}
\end{align}
where $S = \int_0^\beta d \compd
\int d^2 x \sqrt{g} \ ( \mc{L}_{\text{mat}} + \mc{L}_{U(1)} + \mc{L}_{\text{top}})$ and
\begin{align}
&
\mc{L}_\text{mat}
=
-i  ( \opsi \g^\mu [\nabla_\mu - i \ch A_\mu^{(\mc{P})} ] \psi)
 +i \ch( \opsi \si ^{(\mc{P})} \psi) 
- i \ch\ophi  (\olam ^{(\mc{P})} \psi)
 - \frac{i(2\Delta -1)}{2} ( \opsi \g_\compd \psi)  
  +\oF F
+ i \ch(  \opsi \lam^{(\mc{P})})  \phi 
\notag \\ & \qquad \quad
+    |( \pa_\mu - i \ch A_\mu^{(\mc{P})}) \ophi |^2
+   \ophi (\ch \si^{(\mc{P})})^2 \phi  
 + i \ch \ophi D^{(\mc{P})}  \phi  \
- (2\Delta -1)    \ophi [\pa_\compd - i \ch A_\compd^{(\mc{P})}] \phi  
- {\Delta(\Delta -1)}  \ophi \phi ,
%+ \frac{\Delta}{2}  \ophi \phi,
\notag
\\
&\mc{L}_{U(1)} =
 +\frac{1}{2}    (F_{\mu \nu}^{(\mc{P})})^2
 + (D^{(\mc{P})})^2
 +(\pa_\mu \si^{(\mc{P})}) ^2
 + %\frac{1}{f}  
 \e^{\compd  \mu \nu} \si^{(\mc{P})}  F_{\mu \nu} ^{(\mc{P})}
 + %\frac{1}{f^2}  
 (\si^{(\mc{P})})^2 
 %\notag \\
 %&
+ i \olam^{(\mc{P})} \g^\mu \nabla_\mu \lam^{(\mc{P})}
%- i \olam [ \lam , \si]
- \frac{i}{2} \olam^{(\mc{P})} \g_\compd \lam ^{(\mc{P})},
\notag
%\label{YM}
\\
&\mc{L}_{\text{top}} =
\e^{\mu \nu \rho}
F_{\mu \nu}
^{(\mc{P})}
a_\rho(s, \theta_\pm)
%)
%- \olam^{(\mc{P})} \lam^{(\mc{CP})}
%- \olam^{(\mc{CP})} \lam^{(\mc{P})} 
- 2 s D^{(\mc{P})},
%+ 2 D^{(\mc{CP})} \sigma^{(\mc{P})} 
\notag
\end{align}
We write here the \textit{auxiliary fields} $F, \oF, D^{(\mc{P})}$, and after integrating out them, we get a nontrivial potential for $\phi$.
The $\Delta$ is an arbitrary parameter called \textit{R-charge} for the matter multiplet.
In the final line $\mathcal{L}_{\text{top}}$, we use the following configuration parametrized by $(s,\theta_\pm)$
\begin{align}
a(s, \theta_\pm)=s (\cos \theta -1) d\compt  + \frac{\theta_\pm}{2 \pi} d \compd,
\quad
(\compe, \compt) \text{: $\mathbb{RP}^2$ coordinates.}
\end{align}

%%%%%%%%%%%%%%%%%%%%%%%%%%%%
\paragraph{Recipe for the index from localization computation}
Thanks to supersymmetric localization argument, it turns out that the path integral can be calculated \textit{exactly} by considering up to one loop.
More precisely speaking, the Gaussian calculation is performed around configurations giving $S=0$, and all of these contributions should be summed (or integrated) over.
The result is summarized in section 2.2 of \cite{Tanaka:2015pwa} as simple rule.
By using the rule, we can calculate the path integral weighted by the theory characterized by a quiver diagram.
The rule is composed by following steps:
\begin{itemize}
\item
If there is node with rigid line, just replace it as sum of possible dotted objects
\end{itemize}
\begin{align}
& \qquad \qquad 
\Bigg(
\xymatrix {
\dots
\ar@{->}[r]
&
*++[o][F-]{\text{\scriptsize$\mc{P}$}} 
}
\Bigg)
=
q^{+ \frac{1}{8}} \frac{(q^2 ; q^2)_\infty}{(q ; q^2)_\infty}
\sum_{s^{\pm}_{}=0,1}
\frac{1}{2 \pi}
\int_0^{2\pi} d \theta \
\Bigg(
\xymatrix {
\dots
\ar@{.>}[r]
&
*++[o][F.]{\text{\scriptsize$s^\pm, \theta$}} 
}
\Bigg)
,
\label{sum1}
\\ & \qquad \qquad 
\Bigg(
\xymatrix {
\dots
\ar@{->}[r]<1mm>
\ar@{<-}[r]<-1mm>
&
*++[o][F-]{\text{\tiny$\mc{CP}$}} 
}
\Bigg)
=
q^{- \frac{1}{8}} \frac{(q ; q^2)_\infty}{(q^2 ; q^2)_\infty}
\sum_{s \in  \mathbb{Z}}
\frac{1}{2}
\sum_{\theta_\pm = 0,\pi} 
\Bigg(
\xymatrix {
\dots
\ar@{.>}[r]<1mm>
\ar@{<.}[r]<-1mm>
&
*++[o][F.]{\text{\scriptsize$s , \theta_\pm$}} 
}
\Bigg)
,
\label{sum2}
\end{align}
\begin{itemize}
\item
Then, the resultant quiver diagram is represented by only dotted lines and dotted circler nodes. We regard it as the integrand or summand in \eqref{sum1} or \eqref{sum2} by using the following rules\footnote{We correct the formula \eqref{d1loop} in the previous version by adding the exponential prefactor.}
\end{itemize}
%
%\vspace{-1cm}
\begin{align}
&%\qquad
\xymatrix{
*++[o][F.]{\text{\scriptsize$s^\pm_1 , \theta_1$}} 
&
*+[F-]{\Phi}
\ar@{.>}[l]^{\underbrace{}_{\ch_1}} 
\ar@{.>}[r]^{}_{\underbrace{}_{\ch_2}} 
&
*++[o][F.]{\text{\scriptsize$s^\pm_2 , \theta_2$}} 
}
=
\left\{ \begin{array}{ll}
\Big(
q^{ \frac{\Delta - 1}{8}} e^{ \frac{i (\ch_1 \theta_1+\ch_2 \theta_2)}{4} } %\alpha^{ \frac{\fl}{4} }
\Big)^{+1}
	\frac{( e^{- i (\ch_1 \theta_1 + \ch_2 \theta_2)} %\alpha^{- \fl} 
	q^{\frac{(2 - \Delta)}{2}} ; q^{2} )_{\infty}}{( e^{+ i (\ch_1 \theta_1 + \ch_2 \theta_2)} %\alpha^{+ \fl}
	 q^{\frac{(0+\Delta)}{2}} ; q^{2} )_{\infty}} 
	& \text{for} \  (-1)^{\ch_1 s^\pm_1 + \ch_2 s^\pm_2} = +1, \\[.5em]
\Big(
q^{ \frac{\Delta - 1}{8}} e^{ \frac{i (\ch_1 \theta_1 + \ch_2 \theta_2)}{4} } %\alpha^{ \frac{\fl}{4} }
\Big)^{-1}
	\frac{( e^{- i (\ch_1 \theta_2 + \ch_2 \theta_2)} %\alpha^{- \fl} 
	q^{\frac{(4 - \Delta)}{2}} ; q^{2} )_{\infty}}{( e^{+ i (\ch_1 \theta_1 + \ch_2 \theta_2)} %\alpha^{+ \fl}
	 q^{\frac{(2+\Delta)}{2}} ; q^{2} )_{\infty}} & 
	\text{for}\ (-1)^{\ch_1 s^\pm_1 + \ch_2 s^\pm_2} = -1, \\
\end{array} \right.
\label{s1loop}
\\
&\xymatrix{
*++[o][F.]{\text{\scriptsize$s^\pm_1 , \theta_1$}} 
\ar@{<.}[r]<1mm>
\ar@{<.}[r]<-1mm>_{\underbrace{}_{\ch_1}}
&
*+[F-]{^{\Phi_1}_{\Phi_2}}
%\ar@{=}[r]^<{\quad \ (\pm)}_{\ch}
\ar@{.>}[r]<1mm>
\ar@{<.}[r]<-1mm>_{\underbrace{}_{\ch_2}}
&
*++[o][F.]{\text{\scriptsize$s_2 , \theta{_2}{_\pm}$}} 
}
=
e^{i \ch_1 \ch_2 s_{1}^{\pm} \theta{_2}{_\pm}}
 \Big(  q^{\frac{(1-\Delta)}{2}} e^{- i \ch_1 \theta_1} %\alpha^{- \frac{\fl}{2}}  
\Big)^{|\ch_2 s_2|}
\frac{\Big(
%(\pm 1)^{\ch_2}
e^{-i ( \ch_1 \theta{_1} +\ch_2 \theta{_2}{_\pm}) }  %\alpha^{-\fl} 
%e^{-i } 
q^{|\ch_2 s_2|+\frac{(2-\Delta)}{2}} ; q \Big)_\infty}
{\Big(
%(\pm1)^{\ch_2}
e^{+i (\ch_1 \theta{_1}+\ch_2 \theta{_2}{_\pm}  )}  %\alpha^{+\fl} 
%e^{+i \ch_2 \theta{_2}{_\pm}} 
q^{|\ch_2 s_2|+\frac{(0+\Delta)}{2}} ; q\Big)_\infty},
\label{d1loop}
\\
&\xymatrix{
*++[o][F.]{\text{\scriptsize$s_1 , \theta{_1}{_\pm}$}} 
\ar@{.>}[r]<1mm>
\ar@{<.}[r]<-1mm>_{\underbrace{}_{\ch_1}}
&
*+[F-]{^{\Phi_1}_{\Phi_2}}
%\ar@{=}[r]^<{\quad \ (\pm)}_{\ch}
\ar@{.>}[r]<1mm>
\ar@{<.}[r]<-1mm>_{\underbrace{}_{\ch_2}}
&
*++[o][F.]{\text{\scriptsize$s_2 , \theta{_2}{_\pm}$}} 
}=
 \Big(  q^{\frac{1-\Delta}{2}} %e^{- i \ch_1 \theta_1} %\alpha^{- \frac{\fl}{2}}  
\Big)^{| (\ch_1 s_1+ \ch_2 s_2)|}
\frac{\Big(
%(\pm 1)^{\ch_2}
e^{-i  (\ch_1 \theta{_1}{_\pm} + \ch_2 \theta{_2}{_\pm}) }  %\alpha^{-\fl} 
%e^{-i } 
q^{ | (\ch_1 s_1 + \ch_2 s_2)|+\frac{(2-\Delta)}{2}} ; q \Big)_\infty}
{\Big(
%(\pm1)^{\ch_2}
e^{+i (\ch_1 \theta{_1}{_\pm}+ \ch_2 \theta{_2}{_\pm}  )}  %\alpha^{+\fl} 
%e^{+i \ch_2 \theta{_2}{_\pm}} 
q^{ |(\ch_1 s_1+ \ch_2 s_2)|+\frac{(0+\Delta)}{2}} ; q\Big)_\infty},
\label{d1loop2}
\\ & \
\xymatrix{
*++[o][F.]{\text{\scriptsize$s^\pm, \theta$}}
\ar @{.}[r]^{\text{BF}}
&
*++[o][F.]{\text{\scriptsize$s, \theta_\pm$}}
}=
e^{i s^\pm \theta_\pm   }
e^{i s \theta}
,
\qquad
\xymatrix{
*++[o][F.]{\text{\scriptsize$s^\pm, \theta$}}
\ar @{.}[r]^{-\text{BF}}
&
*++[o][F.]{\text{\scriptsize$s, \theta_\pm$}}
}=
e^{-i s^\pm \theta_\pm   }
e^{-i s \theta}
.
\label{BFvalue}
\end{align}

\vspace{.5cm}
We write here doubly charged matter multiplets in \eqref{s1loop}, \eqref{d1loop} and \eqref{d1loop2}, but if one want to consider singly charged matter, the formula is obtained just by turning off one of charges $\ch_1$ or $\ch_2$.
In addition, we can couple matter rectangular node to circular nodes more than three, in this case, the formula is generalized naturally just by replacing
\begin{align}
&\ch s, \ch s^\pm \to \sum_{i} \ch_i s_i,  \sum_{i} \ch_i s_i^\pm,
\notag
\qquad
\ch \theta, \ch \theta_\pm \to \sum_{i} \ch_i \theta_i,  \sum_{i} \ch_i \theta{_i}{_\pm},
\notag
\end{align}
in all expression.
%%%%%%%%%%%%%%%%%%%%%%%%%%%%%%%%%%%%

%%%

%%%%%%%%%%%%%%%%%%
%%%%%%%%%%%%%%%%%%
\include{common_supp_files/sonota/sections_end}
%!TEX encoding = JISJapanese

\include{common_supp_files/sonota/begin2}

%%%%%%%%%%%% Section B %%%%%%%%%%%%%%%%%%%%%%%%%%%%%%%%%%%%%%%
%\section{Mathematical preliminaries}\label{maths}
\subsection{S-operation}\label{cal}

\paragraph{Interpretation for background vector multiplets}
For later use, let us remind the physical meaning of background vector multiplets, e.g. \eqref{example1} or \eqref{example2}.
Basically, turning on $s^\pm$ or $s$ corresponds to changing the Hilbert space because turning on the flux generates Landau levels in the sense of quantum mechanics.
On the other hand, turning on $\theta$ or $\theta_\pm$ is equivalent to consider twisted boundary condition\footnote{
This is true for background vector multiplet coupled to matter multiplets. 
For background vector multiplet coupled to dynamical vector multiplets via topological coupling, one should regard the dynamical vector multiplet as matter multiplet by using dual picture through $*F = d \rho$ where $\rho$ is real scalar.
}:
\begin{align}
&\Bigg(
\xymatrix {
\dots
\ar@{->}[r]
&
*++[o][F.]{\text{\scriptsize$s^\pm , \theta$}} 
}
\Bigg)
\sim
\text{Tr}_{\mc{H}^{\mathbb{RP}^2}_{s^\pm} }
\Big[
(-1)^{\hat{F}}
q^{\frac{1}{2} (\hat{H}+\hat{j}_3)}
e^{-i \theta}
\Big]
:=
\mc{I}^{\mathbb{RP}^2} (q; s^\pm, e^{i\theta})
,
\label{indexP}
\\
&
\Bigg(
\xymatrix {
\dots
\ar@{->}[r]<1mm>
\ar@{<-}[r]<-1mm>
&
*++[o][F.]{\text{\scriptsize$s , \theta_\pm$}} 
}
\Bigg)
\sim
\text{Tr}_{\mc{H}^{\mathbb{RP}^2}_{s} }
\Big[
(-1)^{\hat{F}}
q^{\frac{1}{2} (\hat{H}+\hat{j}_3)}
e^{-i \theta_\pm}
\Big]
:=
\mc{I}^{\mathbb{RP}^2} (q; s, e^{i\theta_\pm}).
\label{indexCP}
\end{align}

\paragraph{S-operation}
Usually, S-operation was defined by using topological BF term as
\begin{align}
\Bigg(
\xymatrix {
\dots
\ar@{->}[r]
&
*+++[o][F.]{} 
}
\Bigg)
%\notag \\
%
%\\&
\quad
\overset{S^{\pm 1}}{\longrightarrow}
\quad
\Bigg(
\xymatrix {
\dots
\ar@{->}[r]
&
*+++[o][F-]{} 
\ar@{.}[r]^{\pm \text{BF}}
&
*+++[o][F.]{} 
}
\Bigg)
.
\label{defS}
\end{align}
However, as we emphasized, there are two types of vector multiplets in our case, and the BF term is only possible between different types of nodes \eqref{Topqui}.
It means that the type of external vector multiplet in \eqref{defS} should be exchanged under the S-operation:
\begin{align}
&\text{$\mc{P}$-type background $\overset{S^{\pm1}}{\longrightarrow}$ $\mc{CP}$-type background}
\notag \\ & \qquad \qquad\qquad
\Bigg(
\xymatrix {
\dots
\ar@{->}[r]
&
*++[o][F.]{\text{\scriptsize$s^\pm , \theta$}} 
}
\Bigg)
%\notag \\
%
%\\&
\quad
\overset{S^{\pm 1}}{\longrightarrow}
\quad
\Bigg(
\xymatrix {
\dots
\ar@{->}[r]
&
*++[o][F-]{\text{\scriptsize$\mc{P}$}} 
\ar@{.}[r]^{\pm \text{BF}}
&
*++[o][F.]{\text{\scriptsize$s , \theta{_\pm}$}} 
}
\Bigg)
\\
&\text{$\mc{CP}$-type background $\overset{S^{\pm1}}{\longrightarrow}$ $\mc{P}$-type background}
\notag \\ & \qquad \qquad\qquad
\Bigg(
\xymatrix {
\dots
\ar@{->}[r]<1mm>
\ar@{<-}[r]<-1mm>
&
*++[o][F.]{\text{\scriptsize$s , \theta_\pm$}} 
}
\Bigg)
%\notag \\
%
%\\&
\quad
\overset{S^{\pm1}}{\longrightarrow}
\quad
\Bigg(
\xymatrix {
\dots
\ar@{->}[r]<1mm>
\ar@{<-}[r]<-1mm>
&
*++[o][F-]{\text{\scriptsize$\mc{CP}$}} 
\ar@{.}[r]^{\pm \text{BF}}
&
*++[o][F.]{\text{\scriptsize$s^\pm , \theta$}} 
}
\Bigg)
\end{align}
%%%
Then, we can observe the expected characters, $S S^{-1} =1, S^2 = -1$ by utilizing the following formulas:
\begin{align}
& \frac{1}{2\pi} \int_0^{2\pi} d\theta \ e^{i s \theta} = \delta_{s,0},
%\qquad
&&\sum_{s \in \mathbb{Z} } e^{i s \theta} = 2 \pi \delta (\theta), \label{mod2deltafn} \\
&\frac{1}{2} \sum_{\theta_\pm =0,\pi } e^{i s^\pm \theta_\pm} =  \delta_{s^\pm ,0}^\text{mod 2},
%\qquad
&&\sum_{s^\pm = 0, 1 } e^{i s^\pm \theta_\pm} = 2 \delta^\text{mod 2} (\theta_\pm), \label{mod2Kdelta}
\end{align}
where we define mod 2 delta functions:
\begin{align}
\delta_{s^\pm ,0}^\text{mod 2}
=
\left\{ \begin{array}{ll}
1 & \text{for } s^\pm \equiv 0 \ (\text{mod }2) \\
0 & \text{for } s^\pm \equiv 1 \ (\text{mod }2) \\
\end{array} \right.
,
\quad
\delta^\text{mod 2} (\theta_\pm)
=
\left\{ \begin{array}{ll}
1 & \text{for } \theta_\pm/\pi  \equiv 0 \ (\text{mod }2) \\
0 & \text{for } \theta_\pm/\pi \equiv 1 \ (\text{mod }2) \\
\end{array} \right.
\end{align}

%%%%%%%%%%%%%%%%%%%%%%%%%%%%%%%%%%%%%%%%%%%%%%%%%%%%%%%%
\include{common_supp_files/sonota/sections_end}

%!TEX encoding = JISJapanese

\include{common_supp_files/sonota/begin2}

%%%%%%%%%%%% Section 3 %%%%%%%%%%%%%%%%%%%%%%%%%%%%%%%%%%%%%%%

\section{Mirror symmetry via loop operators}\label{Mirrorloops}
Once we regard our index in path integral form like in \eqref{pi},
we can insert \text{operator} $\mc{O}$ like
\begin{align}
%\hat{\mc{O}} \cdot \mc{I}^{\mathbb{RP}^2} (q)
%=
\int_{\mathbb{RP}^2 \times \mathbb{S}^1}
\mc{D} \Phi
\mc{D} V
\
\mc{O}
e^{-S}
.
\end{align}
In order to apply the supersymmetric localization technique to calculate it, the operator should preserve part of the supersymmetry, called BPS condition. 
\subsection{BPS loop operators}
We can find such \text{BPS Wilson loop} which wraps $\mathbb{S}^1$-direction standing at north pole = south pole of $\mathbb{RP}^2$.
To define Wilson loop, gauge field is necessary.
In our case, vector multiplet is necessary.
As we noted, there are two types of vector multiplet.
We find the following operators are BPS in each type
\begin{align}
&\mc{P}\text{-type BPS Wilson loop:}
\quad
W_e^{(\mc{P})} = e^{2i e \int_0^{2\pi} d \compd A_\compd |_{\compe=0,\pi}},
\\
&\mc{CP}\text{-type BPS Wilson loop:}
\ \
W_e^{(\mc{CP})} = e^{- i e \int_0^{2\pi} d \compd \ \si |_{\compe=0,\pi}}.
\end{align}
%%%%
On the other hand, one can define \textit{disorder} operator called Vortex loop operator by considering singular boundary condition along the loop.
On $\mathbb{S}^3$ or $\mathbb{S}^2 \times \mathbb{S}^1$, these supersymmetric operators are defined in somewhat direct way \cite{Drukker:2012sr}, but one can utilize S-operation to define them as done on $\mathbb{S}^3$ by \cite{Kapustin:2012iw}.
We have already observed the S-operation works well in our case, so we use this formula to define vortex loops for two-types:
\begin{align}
&\mc{P}\text{-type BPS Vortex loop:}
\quad
V_m^{(\mc{P})} \overset{S}{\leftarrow} W_m^{(\mc{CP})} ,
\\
&\mc{CP}\text{-type BPS Vortex loop:}
\ \
V_m^{(\mc{CP})} \overset{S}{\leftarrow}  W_m^{(\mc{P})} .
\end{align}
Now, from the point of view that S-operation is ``particle-vortex duality", it is understood as $\mc{P}$-type vortex = $\mc{CP}$-type particle, and $\mc{CP}$-type vortex = $\mc{P}$-type particle.
See Appendix \ref{Loops} for more detail.

%%%%%%%%%%
\subsection{Loop operator effects on localization computation}
First, let us define quiver notation for inserting the above loop operators.
We just represent insertion of loop operator $\mathcal{O} = W_e , V_m$ associated symmetry by just adding appropriate subscript to the circular node:
\begin{align}
\hat{\mathcal{O}} \cdot \mc{I}^{\mathbb{RP}^2} (q)
:=
\int \dots \mc{D}V \ \mathcal{O} (V) e^{-S}
=
\Bigg(
\xymatrix {
\dots
\ar@{->}[r]_>{\qquad \quad \ \ \mathcal{O}}
&
*+++[o][F-]{} 
}
\Bigg),
\end{align}
then, the recipe for the exact result is as follows:

\begin{itemize}
\item
1. Follow the previous recipe in \eqref{sum1} - \eqref{BFvalue}
\item
2. For the subscript $\mathcal{O} = W_e$ or $V_m$,  follow the rules:
\end{itemize}
\begin{align}
&
\Bigg(
\xymatrix {
\dots
\ar@{->}[r]_>{\qquad \qquad \qquad \ W_e}
&
*++[o][F.]{\text{\scriptsize$s^\pm , \theta$}} 
}
\Bigg)
 =
e^{2 i e \theta} 
\Bigg(
\xymatrix {
\dots
\ar@{->}[r]
&
*++[o][F.]{\text{\scriptsize$s^\pm , \theta$}} 
}
\Bigg)
,
\
\Bigg(
\xymatrix {
\dots
\ar@{->}[r]_>{\qquad \qquad \qquad \ V_m}
&
*++[o][F.]{\text{\scriptsize$s^\pm , \theta$}} 
}
\Bigg) =
e^{-2  \pi m \frac{\partial}{\partial \theta} } 
\Bigg(
\xymatrix {
\dots
\ar@{->}[r]
&
*++[o][F.]{\text{\scriptsize$s^\pm , \theta$}} 
}
\Bigg),
\label{VP}
\\
&%W_e \cdot \mc{I}_{\mc{CP}} (B, \theta^\pm) 
\Bigg(
\xymatrix {
\dots
\ar@{->}[r]<1mm>
\ar@{<-}[r]<-1mm>_>{\qquad \qquad \qquad \ W_e}
&
*++[o][F.]{\text{\scriptsize$s , \theta_\pm$}} 
}
\Bigg)
=
e^{2 \pi i e s} 
\Bigg(
\xymatrix {
\dots
\ar@{->}[r]<1mm>
\ar@{<-}[r]<-1mm>
&
*++[o][F.]{\text{\scriptsize$s , \theta_\pm$}} 
}
\Bigg)
,
\
\Bigg(
\xymatrix {
\dots
\ar@{->}[r]<1mm>
\ar@{<-}[r]<-1mm>_>{\qquad \qquad \qquad \ V_m}
&
*++[o][F.]{\text{\scriptsize$s , \theta_\pm$}} 
}
\Bigg)=
e^{- 2 m \frac{\partial }{\partial s} }
\Bigg(
\xymatrix {
\dots
\ar@{->}[r]<1mm>
\ar@{<-}[r]<-1mm>
&
*++[o][F.]{\text{\scriptsize$s , \theta_\pm$}} 
}
\Bigg).
\label{VCP}
\end{align}
%%%%
\paragraph{Generalized index and its loop insertions}
We can regard the above formulas in \eqref{VP} and \eqref{VCP} in the context of generalized indices \eqref{indexP} or \eqref{indexCP}.
In this case, %we use 
\begin{align}
&\text{wilson loop: $w_e$,}
\quad
\text{vortex loop: $v_m$}
\end{align}
are associated by the external fields, and we get
\begin{align}
&\hat{w}_e \cdot  \mc{I}^{\mathbb{RP}^2} (q;\ s^\pm , \theta)
=
e^{2i e \theta} \mc{I}^{\mathbb{RP}^2} (q;\ s^\pm , \theta),
\quad
\hat{v}_m \cdot \mc{I}^{\mathbb{RP}^2} (q;\ s^\pm , \theta)
=
\mc{I}^{\mathbb{RP}^2} (q;\ s^\pm , \theta -2\pi m),
\\%%%%%%%%%%%%
&\hat{w}_e \cdot \mc{I}^{\mathbb{RP}^2} (q;\ s , \theta_\pm)
=
e^{2 \pi i e s} \mc{I}^{\mathbb{RP}^2} (q;\ s , \theta_\pm),
\quad
\hat{v}_m \cdot \mc{I}^{\mathbb{RP}^2} (q;\ s , \theta_\pm)
=
\mc{I}^{\mathbb{RP}^2} (q;\ s -2m , \theta_\pm),
\end{align}
One can find the Vortex operators act on the indices as finite differential operators.
%
%\paragraph{Loop operators associated with gauge symmetries}
We can regard the gauging procedure as the S-transform itself, i.e.
\begin{align}
%\big( 
\hat{W}_e %\cdot \mc{I}^{\mathbb{RP}^2} \big) (q; s^\pm , \theta)
=
%\big(
\hat{S} \hat{w}_e 
%\cdot  \mc{I}^{\mathbb{RP}^2} \big)  (q; s^\pm , \theta)
,
\quad
\hat{V}_m
=
\hat{S} \hat{v}_{m}.
\end{align}
%%%%%%%%
Therefore, we get
\begin{align}
\hat{W}_e
=
\hat{S} \hat{w}_e  \hat{S}^{-1} \hat{S}
=
\hat{v}_{-e} \hat{S}
,
\quad
\hat{V}_m
=
\hat{S} \hat{v}_m  \hat{S}^{-1} \hat{S}
=
\hat{w}_{-m} \hat{S}.
\label{gauge-global}
\end{align}
Note that this equalities mean the statements in \eqref{mirrorloopP1} and \eqref{mirrorloopCP1} because S-transformation exchanges the $\mc{P}$ and $\mc{CP}$.

\subsection{Mirror symmetry with loop operators}
The expectation values of loop operators associated to $U(1)_A$ symmetry should be in agreement if the original indices are identical.
We can find somewhat non-trivial statement in \cite{Drukker:2012sr, Assel:2015oxa} that 
\begin{align}
&\bullet \ \text{Gauge Wilson loop for SQED} \Leftrightarrow \text{Flavor $U(1)_V$ Vortex loop for XYZ.}\label{gaugeW1}
\\
&\bullet \ \text{Gauge Vortex loop for SQED} \Leftrightarrow \text{Flavor $U(1)_V$ Wilson loop for XYZ.} \label{gaugeV1}
\end{align}
These can be easily checked in our case. Because of the algebraic equalities \eqref{gauge-global}, we get
%It is easily checked in our case.
%Because of the algebraic equalities in \eqref{gauge-global}, we easily get that
\begin{align}
&\bullet \ \text{Gauge Wilson loop for SQED}=
\text{Topological $U(1)_J$ Vortex loop for SQED.}
\\
&\bullet \ \text{Gauge Vortex loop for SQED}=
\text{Topological $U(1)_J$ Wilson loop for SQED.}
\end{align}
In addition, we already know that topological U$(1)_{J}$ in the SQED corresponds to flavor U$(1)_{V}$ in the XYZ model under the mirror map. As a consequence, the dualities of the loop operator insertions are claimed between for the gauge symmetry and for the global symmetry as \eqref{gaugeW1} and \eqref{gaugeV1}.

%%%%%%%%%%%%%%%%%%%%%%%%%%%%%%%%%%%%%%%%%%%%%%%%%%%%%%%%
\include{common_supp_files/sonota/sections_end}
%!TEX encoding = JISJapanese

\include{common_supp_files/sonota/begin2}

%%%%%%%%%%%% Section 4 %%%%%%%%%%%%%%%%%%%%%%%%%%%%%%%%%%%%%%%
\section{Multi-flavor mirror symmetry} \label{Multiflavor}

%%%%%%%%%%%%%%%%%% section 4.1 %%%%%%%%%%%%%%%%%%%%%%%%%%%%%%%%
\subsection{From single-flavor to multi-flavor mirror symmetry \label{Prescription}}
Abelian 3d ${\cal N} = 2$ mirror symmetry \cite{Intriligator:1996ex, Aharony:1997bx, Kapustin:1999ha, Tong:2000ky} with a single matter is the duality between the SQED and the XYZ model, namely, they flow to the same fixed point in IR. The moduli parameters in the SQED, the combination of the scalar in the vector multiplet and the dual photon
%\footnote{The dual photon $\rho$ is defined by
%\begin{align*}
%{\rm d} \rho = \ast F,
%\end{align*}
%where $F$ is a field strength of the dynamical gauge field.}
$e^{\pm ( \sigma + i \rho )}$ for Coulomb branch, and the product of matters $\tilde{Q} Q$ for Higgs branch, correctly match $( X, Y )$ and $Z$ which are matter multiplets in the XYZ model, respectively. 
Furthermore, there are two global symmetries U$(1)_{J}$ (as a topological one) and U$(1)_{A}$ in addition to an R-symmetry in the SQED, and the XYZ model has exactly corresponding flavor symmetries U$(1)_{V}$ and U$(1)_{A}$ acting on the matter fields as for usual dualities. For the neat review of this simplest version of mirror symmetry and charge assignments on the fields, see \cite{Kapustin:2011jm, Tanaka:2014oda}. This symmetry can be directly checked as the equality of the superconformal indices computed by the localization \cite{Krattenthaler:2011da, Kapustin:2011jm},
\begin{align} \label{N=2mirror}
\mathcal{I}_{\text{{\tiny SQED}}}^{\mathcal{N} = 2}
=
\mathcal{I}_{\text{{\tiny XYZ}}}^{\mathcal{N} = 2},
\end{align}
for the SQED $\mathcal{I}_{\text{{\tiny SQED}}}^{\mathcal{N} = 2}$ and the XYZ model $\mathcal{I}_{\text{{\tiny XYZ}}}^{\mathcal{N} = 2}$ (from now on, we omit the superscript $\mathbb{RP}^2$ on the index).

%We can immediately construct the ${\cal N} = 4$ version of mirror symmetry. On the SQED side, this is done only by introducing an adjiont chiral multiplet $\tilde{S}$ coupled to $\tilde{Q} Q$. Correspondingly, in the XYZ model, coupling an additional chiral multiplet $\tilde{Z}$ to $Z$ makes both massive, then they are integrated out, which leads to the free theory only with $X$ and $Y$.
We can enlarge supersymmetry to $\mathcal{N} = 4$ \cite{Intriligator:1996ex, Aharony:1997bx, Kapustin:1999ha} by adding an adjoint chiral multiplet $\tilde{S}$ to the $\mathcal{N} = 2$ SQED in such a way that it couples with the matters through a superpotential $\tilde{Q} \tilde{S} Q$. Correspondingly, in the XYZ model, we need to couple a chiral multiplet $\tilde{Z}$ to $Z$, which is equivalent to introducing the mass term $\tilde{Z} Z$ for them. Thus, they are integrated out, and the resultant matter theory becomes the free theory only with $X$ and $Y$. In the language of the index which is known to be a protected quantity along the RG flow, those processes can be done simply by multiplying the inverse of the one-loop determinant $\mathcal{Z}_{\text{{\tiny 1-loop}}}^{Z}$ for $Z$, or equivalently, the one-loop determinant $\mathcal{Z}_{\text{{\tiny 1-loop}}}^{\tilde{Z}}$ for $\tilde{Z}$, by both sides of the equality \eqref{N=2mirror}. %Schematically, we have
%\begin{align} \label{N=4mirror1}
%\mathcal{I}_{\text{{\tiny SQED}}}^{\mathcal{N} = 4}
%=
%\mathcal{I}_{\text{{\tiny XY}}}^{\mathcal{N} = 4}
%\Leftrightarrow
%(
%\mathcal{Z}_{\text{{\tiny 1-loop}}}^{Z}
%)^{- 1}
%\mathcal{I}_{\text{{\tiny SQED}}}^{\mathcal{N} = 2}
%=
%(
%\mathcal{Z}_{\text{{\tiny 1-loop}}}^{Z}
%)^{- 1}
%\mathcal{I}_{\text{{\tiny XYZ}}}^{\mathcal{N} = 2},
%\end{align}
This is just the observation that integrating out $Z$ in the XYZ model is to simply remove its one-loop determinant from the index, and, on the other hand, the new adjoint field $\tilde{S}$ in the SQED has charges under the global symmetries corresponding to these of $\tilde{Z}$ which can be read off from the superpotential. Thus, the contribution $\mathcal{Z}_{\text{{\tiny 1-loop}}}^{\tilde{S}}$ for $\tilde{S}$ to the index can be viewed as the one $\mathcal{Z}_{\text{{\tiny 1-loop}}}^{\tilde{Z}}$ for $\tilde{Z}$ under mirror symmetry, and then schematically, $\mathcal{N} = 4$ mirror symmetry should be written as
\begin{align} \label{N=4mirror2}
\mathcal{I}_{\text{{\tiny SQED}}}^{\mathcal{N} = 4}
=
\mathcal{I}_{\text{free}}^{\mathcal{N} = 4}
\hspace{1em}
\Leftrightarrow
\hspace{1em}
\mathcal{Z}_{\text{{\tiny 1-loop}}}^{\tilde{S}}
\mathcal{I}_{\text{{\tiny SQED}}}^{\mathcal{N} = 2}
=
\mathcal{Z}_{\text{{\tiny 1-loop}}}^{\tilde{Z}}
\mathcal{I}_{\text{{\tiny XYZ}}}^{\mathcal{N} = 2},
\end{align}
where $\mathcal{I}_{\text{{\tiny SQED}}}^{\mathcal{N} = 4}$ and $\mathcal{I}_{\text{free}}^{\mathcal{N} = 4}$ are the indices for the $\mathcal{N} = 4$ SQED and for the free matter theory, respectively, and
\begin{align}
\mathcal{Z}_{\text{{\tiny 1-loop}}}^{\tilde{S}}
=
\mathcal{Z}_{\text{{\tiny 1-loop}}}^{\tilde{Z}}
=
(
\mathcal{Z}_{\text{{\tiny 1-loop}}}^{Z}
)^{- 1}
\end{align}
under the suitable identification of parameters. The index formula \eqref{N=4mirror2} of ${\cal N} = 4$ mirror symmetry with a single flavor is a key to demonstrate ${\cal N} = 2$ mirror symmetry with many flavors. We will show these explicit forms depending on the parity conditions in the next subsection.

Now, we aim to briefly review the statement of ${\cal N} = 2$ multi-flavor mirror symmetry. The one side of the duality is the SQED with $N_{f}$ a general number of matters. The dual side is the quiver theory with an U$(1)^{N_{f} - 1}$ gauge symmetry, and there are an adjoint matter associated with each gauge node, bifundamental matters connecting two neighboring gauge nodes, and fundamental matters at the ends of the quiver diagram. The former can be obtained from the theory with $N_{f}$ free matters by gauging the sum of $N_{f}$ U$(1)_{V}$'s equipped with each single-flavor free theory. To construct the correct dual theory, we start with $N_{f}$ sets of the ${\cal N} = 4$ SQED comprised of the ${\cal N} = 2$ SQED and an adjoint matter coupled to the fundamental matter via the superpotential, then gauging the sum of $N_{f}$ U$(1)_{J}$'s. Note that this gauging process is equivalent to ungauging the diagonal U$(1)$ part of the U$(1)^{N_{f}}$ gauge symmetry, which results in the U$(1)^{N_{f} - 1}$ quiver gauge theory with appropriately redefining  the gauge charges of the bifundamental fields. Actually, the above theories which we use as starting points to get multi-flavor mirror symmetry are exactly connected to each other by ${\cal N} = 4$ mirror symmetry: as explained, the $N_{f} = 1$ free matter theory is dual to the $\mathcal{N} = 4$ SQED with a single flavor. Therefore, we take the following strategy \cite{Kapustin:2011jm} based on that fact to build multi-flavor mirror symmetry in terms of the index:
	\begin{enumerate} % Strategy for N=4 mirror sym
	%1
	\item We set the index $\mathcal{I}_{N_{f}}$ as the collection of $N_{f}$ free single-matter theories, that is, $\mathcal{I}_{N_{f}} = \prod_{i = 1}^{N_{f}} \mathcal{I}_{\text{free}}^{\mathcal{N} = 4} ( a_{i} )$ where $a_{i}$ is some parameter of the $i$-th free theory. This clearly enjoys ${\cal N} = 4$ mirror symmetry as
	\begin{align} \label{N=4multimirror}
	\mathcal{I}_{N_{f}}
	=
	\mathcal{I}_{\text{U}(1)^{N_{f}}},
	\end{align}
	where $\mathcal{I}_{\text{U}(1)^{N_{f}}}$ is the index for $N_{f}$ copies of the $\mathcal{N} = 4$ SQED with one flavor, $\mathcal{I}_{\text{U}(1)^{N_{f}}} = \prod_{i = 1}^{N_{f}} \mathcal{I}_{\text{{\tiny SQED}}}^{\mathcal{N} = 4} ( a_{i} )$.
	%2
	\item For $\mathcal{I}_{N_{f}}$ in \eqref{N=4multimirror} which contains global symmetries U$(1)_{V}^{N_{f}}$ and U$(1)_{A}^{N_{f}}$, we gauge the diagonal part of U$(1)_{V}^{N_{f}}$ by taking the suitable combinations of parameters and then obtain the index $\mathcal{I}_{\text{SQED}_{N_{f}}}$ for the SQED with general $N_{f}$ flavors.
	%3
	\item For $\mathcal{I}_{\text{U}(1)^{N_{f}}}$ in \eqref{N=4multimirror} which contains global symmetries U$(1)_{J}^{N_{f}}$ and U$(1)_{A}^{N_{f}}$, we gauge the diagonal part of U$(1)_{J}^{N_{f}}$ by taking the suitable combinations of parameters and adjust gauge parameters to show the correct charges of the bifundamental matters. As a result, we get the index $\mathcal{I}_{\text{U}(1)^{N_{f} - 1}}$ for the U$(1)^{N_{f} - 1}$ quiver gauge theory.
	%4
	\item Finally, by construction, we conclude multi-flavor mirror symmetry as the index equality
	\begin{align}
	\mathcal{I}_{\text{SQED}_{N_{f}}}
	=
	\mathcal{I}_{\text{U}(1)^{N_{f} - 1}}.
	\end{align}
	\end{enumerate}

In the following, we say a $\mathcal{P}$-type or a $\mathcal{CP}$-type U$(1)$ symmetry for short if the background gauge field for that U$(1)$ respects the $\mathcal{P}$-type or the $\mathcal{CP}$-type parity condition, respectively.

%%%%%%%%%%%%%%%%%% section 4.2 %%%%%%%%%%%%%%%%%%%%%%%%%%%%%%%%
\subsection{Single-flavor $\mathcal{N} = 4$ mirror symmetry}
In this subsection, we will give the formulas for $\mathcal{N} = 4$ mirror symmetry which play a central role to derive the multi-flavor version of $\mathcal{N} = 2$ mirror symmetry. As explained, they can be readily derived from $\mathcal{N} = 2$ mirror symmetry whose detailed discussions including the charge assignment under the symmetries and rigorous proof can be seen in \cite{Tanaka:2014oda, Tanaka:2015pwa}. 

There are two types of Abelian mirror symmetry on $\Rp^2 \times \Sp^1$ depending on the case where the dynamical vector multiplet in the SQED is set to be ${\cal P}$-type or ${\cal CP}$-type. We name the former ${\cal P}$-SQED$^{( \pm )}$ and the latter ${\cal CP}$-SQED$^{( \pm )}$. The superscript $( \pm )$ represents a fixed value of the background field coupled to the dynamical one though a BF term.

\paragraph{$\mathcal{N} = 4$ ${\cal P}$-SQED$^{( \pm )}$ vs The free matter theory.} %%%%%%%%%%%%%%
%Let us turn to showing Abelian $\mathcal{N} = 4$ mirror symmetry in terms of the index. As mentioned above, the $\mathcal{N} = 4$ index equalities can easily be obtained; From \eqref{mirrorN2P+} with the quiver \eqref{quiverN2P+}, $\mathcal{N} = 4$ mirror symmetry between ${\cal P}$-SQED$^{( \pm )}$ and the corresponding free theory can be given in an unified form as
% N=4 qiuiver of P-SQED^+- and corresponding XYZ
%\begin{align} \label{quiverN4P}
%% P-SQED^+-
%\begin{xy}
%\ar@{} (0,0) *+[F-]{\text{$Q$}} = "B",
%\ar@{->} "B";(15,0) *++[o][F-]{\text{\scriptsize$\mathcal{P}$}} = "A",
%%\ar@{} (30,9.5) | {\text{\Large \rotatebox{180}{$\circlearrowright$}}}
%\ar@{.>>} (15,10);"A"
%\ar@{->} "A";(30,0) *+[F-]{\text{$\tilde{Q}$}},
%\ar@{<.} (0,3);(15,15) *++[o][F.]{\text{\scriptsize$s_{A}^{+}, \theta_A$}};
%\ar@{<.} (19,12);(30,4),
%\ar@{.}_{\text{BF}} (15,-3);(15,-10);
%\ar@{} (15,-15) *++[o][F.]{\text{\scriptsize$s, \theta_{J \pm}$}}
%\end{xy}
%&&
%\Longleftrightarrow
%&&
%% free theory
%\begin{xy}
%\ar@{} (-20,0) *+[F-]{\text{\scriptsize$\begin{matrix} X & Y \end{matrix}$}};
%\ar@{<.} (-21,3);(-21,10);
%\ar@{<.} (-19,3);(-19,10);
%\ar@{} (-20,15) *++[o][F.]{\text{\scriptsize$\tilde{s}_{A}^{+}, \tilde{\theta}_A$}};
%\ar@{<.} (-21,-3);(-21,-10);
%\ar@{.>} (-19,-3);(-19,-10);
%\ar@{} (-20,-15) *++[o][F.]{\text{\scriptsize$\tilde{s}, \theta_{V \pm}$}},
%%\ar@{.>} (4,12);(20,0) *+[F-]{\text{$Z$}},
%\end{xy}
%\end{align}
% Index for N=4 mirror sym v2
\begin{align}
&
\mathcal{I}_{\text{{\tiny ${\cal P}$-SQED$^{( \pm )}$}}}^{\mathcal{N} = 4}
( q; a; s, e^{i \theta_{J \pm}} ) \notag \\ %1
&=
a^{\frac{1}{4}}
\frac{( q^{2}, a^{- 1} q ; q^{2} )_{\infty}}{( q, a ; q^{2} )_{\infty}}
\sum_{s^{\pm}}
\oint \frac{{\rm d} z}{2 \pi i z}
e^{i s^{\pm} \theta_{J \pm}}
z^{s}
\left\{
a^{- \frac{1}{4} + \frac{s^{\pm}}{2}}
\frac
{( z^{- 1} a^{\frac{1}{2}} q^{\frac{1}{2} + s^{\pm}}, z a^{\frac{1}{2}} q^{\frac{1}{2} + s^{\pm}}; q^{2} )_{\infty}}
{( z a^{- \frac{1}{2}} q^{\frac{1}{2} + s^{\pm}}, z^{- 1} a^{- \frac{1}{2}} q^{\frac{1}{2} + s^{\pm}} ; q^{2} )_{\infty}}
\right\}, \label{mirrorN4P-v2} \\ %2
&
\mathcal{I}_{\text{free}}^{\mathcal{N} = 4}
( q; \tilde{a}; \tilde{s}, e^{i \theta_{V  \pm }} )
=
\left( \tilde{a}^{- \frac{1}{2}} q^{\frac{1}{2}} \right)^{| \tilde{s} |}
\frac
{( e^{i \theta_{V \pm}} \tilde{a}^{- \frac{1}{2}} q^{1 + |\tilde{s}|} ; q )_{\infty}}
{( e^{i \theta_{V \pm}} \tilde{a}^{\frac{1}{2}} q^{|\tilde{s}|} ; q )_{\infty}}. %3
\label{mirrorN4P+v2}
\end{align}
The index formula for $\mathcal{N} = 4$ ${\cal P}$-SQED$^{( \pm )}$ is given by \eqref{mirrorN4P-v2}. The parameter $a := e^{- 2 i \theta_{A}} q^{( 1 - \Delta )}$ includes a Wilson line phase $\theta_{A}$ associated with an axial U$(1)_{A}$ global symmetry, and $( s, \theta_{J \pm} )$ are a flux and a discretized Wilson line for $\mathcal{CP}$-type U$(1)_{J}$. The leftmost $q$-Pochhammer symbol of \eqref{mirrorN4P-v2} corresponds to the contribution for the adjoint matter with charge $- 2$ under U$(1)_{A}$ in addition to the U$(1)$ vector multiplet. \eqref{mirrorN4P+v2} is the index for the dual free theory with $\tilde{a} := e^{+ 2 i \tilde{\theta}_{A}} q^{( 1 - \Delta )}$ where $\tilde{\theta}_{A}$ is a Wilson line for U$(1)_{A}$, and $( \tilde{s}, \theta_{V \pm} )$ are a flux and a discretized Wilson line for $\mathcal{CP}$-type U$(1)_{V}$ in this theory. Single-flavor $\mathcal{N} = 4$ mirror symmetry is indicated by the equality of these indices \eqref{mirrorN4P-v2} and \eqref{mirrorN4P+v2} under the map $a = \tilde{a}$ and $( s, \theta_{J \pm} ) = ( \tilde{s}, \theta_{V \pm} )$.
%The diagram \eqref{quiverN4P} is one of the descriptions of $\mathcal{N} = 4$ mirror symmetry: $\mathcal{N} = 4$ ${\cal P}$-SQED$^{( \pm )}$ on the lhs and the dual free theory on the rhs. The circle arrow attached to the gauge node in \eqref{quiverN4P} represents an adjoint chiral field forming a superpotential with $Q$ and $\tilde{Q}$. In the index formula \eqref{mirrorN4P+v2} for this duality, correspondingly, we put the index for ${\cal P}$-SQED$^{( \pm )}$ on the lhs, where $\sum_{s^{\pm}}$ is the summation over holonomies $s^+ = 0$ and $s^- = 1$, and the one for the dual theory on the rhs. Note that the definition of $a$ ($\tilde{a}$) includes a fugacity for an axial U$(1)_{A}$ global symmetry in each theory such that $a = e^{- 2 i \theta_{A}} q^{( 1 - \Delta )}$ ($\tilde{a} = e^{+ 2 i \tilde{\theta}_{A}} q^{( 1 - \Delta )}$).

\paragraph{$\mathcal{N} = 4$ ${\cal CP}$-SQED$^{( \pm )}$ vs The free matter theory.} %%%%%%%%%%%%%%
%We can get between ${\cal CP}$-SQED$^{( \pm )}$ and the corresponding free theory by rewriting \eqref{mirrorN2CP+} with the quiver \eqref{quiverN2CP+} as
% N=4 qiuiver of CP-SQED^+- and corresponding XYZ
%\begin{align} \label{quiverN4CP}
%% CP-SQED^+-
%\begin{xy}
%\ar@{<.} (11.5,2);(24,13);
%\ar@{<.} (11.5,-1);(25,11);
%\ar@{} (8.5,0) *+[F-]{\text{\scriptsize$\begin{matrix} Q \\ \tilde{Q} \end{matrix}$}},
%\ar@{<-} (11,1.5);(25,1.5);
%\ar@{->} (11,-1.5);(25,-1.5);
%\ar@{} (28.5,0) *++[o][F-]{\text{\scriptsize$\mathcal{CP}$}},
%%\ar@{} (57.5,11) | {\text{\Large \rotatebox{180}{$\circlearrowright$}}}
%\ar@{.>>} (28.5,10);(28.5,3.5)
%\ar@{} (28.5,15) *++[o][F.]{\text{\scriptsize$s_{A}^{+}, \theta_A$}};
%\ar@{.}^{\text{BF}} (28.5,-10);(28.5,-4);
%\ar@{} (28.5,-15) *++[o][F.]{\text{\scriptsize$s^{\pm}, \theta_{J}$}}
%\end{xy}
%\hspace{-1.5em}
%&&
%\Longleftrightarrow
%&&
%\hspace{-1.5em}
%% free theory
%\begin{xy}
%\ar@{} (-30,0) *+[F-]{\text{$X$}};
%(-15,0) *+[F-]{\text{$Y$}} = "A",
%%\ar@{} "A";(30,0) *+[F-]{\text{Z}},
%\ar@{<.} (-30,3);(-15,15) *++[o][F.]{\text{\scriptsize$\tilde{s}_{A}^{+}, \tilde{\theta}_A$}};
%\ar@{<.} (-15,2.5);(-15,10);
%%\ar@{.>>} (19,12);(28,3),
%\ar@{.>} "A";(-15,-10);
%\ar@{<.} (-30,-3);(-19,-11);
%\ar@{} (-15,-15) *++[o][F.]{\text{\scriptsize$\tilde{s}^{\pm}, \theta_{V}$}}
%\end{xy}
%\end{align}
% N=4 mirror sym v2
\begin{align}
&
\mathcal{I}_{\text{{\tiny ${\cal CP}$-SQED$^{( \pm )}$}}}^{\mathcal{N} = 4}
( q; a; s^{\pm}, w ) \notag \\ %1
&=
a^{- \frac{1}{4}}
\frac{( q, a; q^{2} )_{\infty}}{( q^{2}, a^{- 1} q; q^{2} )_{\infty}}
\sum_{s \in \mathbb{Z}}
\frac{1}{2}
\sum_{\theta_{\pm}}
e^{i s^{\pm} \theta_{\pm}}
w^{s}
\left( q^{\frac{1}{2}} a^{- \frac{1}{2}} \right)^{| s |}
\left[
\frac{( e^{i \theta_{\pm}} a^{- \frac{1}{2}} q^{| s | + 1}; q )_{\infty}}{( e^{i \theta_{\pm}} a^{\frac{1}{2}} q^{| s |}; q )_{\infty}}
\right], \label{mirrorN4CP-v2} \\ %2
&
\mathcal{I}_{\text{free}}^{\mathcal{N} = 4}
( q; \tilde{a}; \tilde{s}^{\pm}, \tilde{w} )
=
\tilde{a}^{- \frac{1}{4} + \frac{\tilde{s}^{\pm}}{2}}
\frac
{( \tilde{a}^{\frac{1}{2}} \tilde{w}^{- 1} q^{\frac{1}{2} + \tilde{s}^{\pm}}, \tilde{a}^{\frac{1}{2}} \tilde{w} q^{\frac{1}{2} + \tilde{s}^{\pm}}; q^{2} )_{\infty}}
{( \tilde{a}^{- \frac{1}{2}} \tilde{w} q^{\frac{1}{2} + \tilde{s}^{\pm}}, \tilde{a}^{- \frac{1}{2}} \tilde{w}^{- 1} q^{\frac{1}{2} + \tilde{s}^{\pm}}; q^{2} )_{\infty}}. %3
\label{mirrorN4CP+v2}
\end{align}
The other version of $\mathcal{N} = 4$ mirror symmetry is described by the indices \eqref{mirrorN4CP-v2} for $\mathcal{N} = 4$ ${\cal CP}$-SQED$^{( \pm )}$ and \eqref{mirrorN4CP+v2} for the dual free theory. In the former, the parameter $s$ is a holonomy and $w$ is a fugacity for $\mathcal{P}$-type U$(1)_{J}$, on the other hand, $( \tilde{s}^{\pm}, \tilde{w} )$ are a holonomy and a fugacity for $\mathcal{P}$-type U$(1)_{V}$ in the dual theory. $\mathcal{N} = 4$ mirror symmetry with one flavor is realized as the equality of \eqref{mirrorN4CP-v2} and \eqref{mirrorN4CP+v2} under identifications $a = \tilde{a}$ and $( s^{\pm}, w ) = ( \tilde{s}^{\pm}, \tilde{w} )$. The mathematical proof to show   equivalence of the indices in both cases is done in \cite{Tanaka:2014oda, Tanaka:2015pwa}.
%The diagram \eqref{quiverN4CP} is the other of $\mathcal{N} = 4$ mirror symmetry: $\mathcal{N} = 4$ ${\cal CP}$-SQED$^{( \pm )}$ on the lhs and its dual free theory on the rhs. Similarly, the formula \eqref{mirrorN4CP+v2} consists of the indices for ${\cal CP}$-SQED$^{( \pm )}$ on the lhs, where $\sum_{\theta_{\pm}}$ is the summation over two values $\theta_{+} = 0$ and $\theta_{-} = \pi$, and for the dual theory on the rhs.

We should note that the background gauge holonomy for U$(1)_A$ is fixed to be even in all above index formulae. Actually, we can also consider its odd-holonomy by using the revised index formula \eqref{d1loop}, namely, one more parameter for U$(1)_A$ is brought into the indices of the SQED and the XYZ model. We now skip to show detailed expressions, but the statement of Abelian mirror symmetry on $\Rp^2 \times\Sp^1$ still holds at the level of the index.

%%%%%%%%%%%%%%%%%% section 4.3 %%%%%%%%%%%%%%%%%%%%%%%%%%%%%%%%
\subsection{Multi-flavor $\mathcal{N} = 2$ mirror symmetry}
We will demonstrate systematically the multi-flavor version of Abelian mirror symmetry on $\Rp^2 \times\Sp^1$ following the prescription explained in Section \ref{Prescription}. The starting point is the index $\mathcal{I}_{N_{f}}$ of $N_{f}$ free matter multiplets coupled to the background gauge fields associated with $N_{f}$ U$(1)_{V}$ flavor symmetries which rotate a phase to each matter multiplet. We should notice that even though \textit{a priori} each background field can take freely a $\mathcal{P}$- or a $\mathcal{CP}$-type parity condition, it seems to be only allowed to set all of parity conditions being \textit{either} $\mathcal{P}$- or $\mathcal{CP}$-type because we may gauge the diagonal part of U$(1)_{V}^{N_{f}}$ if all of them are the same. It is apparently unable to construct multi-flavor mirror symmetry with the combination of an arbitrary number of $\mathcal{P}$- and $\mathcal{CP}$-type U$(1)_{V}$'s. This is why, in this paper, we have two kinds of the SQEDs in the context of multi-flavor mirror symmetry on $\Rp^2 \times\Sp^1$, that is, $\mathcal{P}$-SQED$_{N_{f}}$ which is the theory obtained by gauging the sum of $\mathcal{P}$-type U$(1)_{V}^{N_{f}}$ in the $N_{f}$-free matter theory and $\mathcal{CP}$-SQED$_{N_{f}}$ by gauging the sum of $\mathcal{CP}$-type U$(1)_{V}^{N_{f}}$.

\paragraph{$\mathcal{P}$-SQED$_{N_{f}}$ vs $\mathcal{CP}$-{\rm U}$(1)^{N_{f} - 1}$ gauge theory.} %%%%%%%%
We first consider $\mathcal{P}$-SQED$_{N_{f}}$ and its dual theory. We begin with the index $\mathcal{I}_{N_{f}}^{( \mathcal{P} )}$ for the free theory with $N_{f}$ $\mathcal{P}$-type U$(1)_{V}$'s given by
%We begin with indicating multi-flavor mirror symmetry between the SQED$_{N_{f}}$ of the $\mathcal{P}$-type gauge symmetry and the dual quiver theory. To do this, we first consider the index $\mathcal{I}_{N_{f}}^{( \mathcal{P} )}$ for the free theory under $N_{f}$ $\mathcal{P}$-type U$(1)_{V}$'s given by
\begin{align} \label{freeNfP}
\mathcal{I}_{N_{f}}^{( \mathcal{P} )} ( q; \{ a_{i} \}; \{ s_{i}^{\pm} \}, \{ w_{i} \} )
=
\prod_{i = 1}^{N_{f}}
a_{i}^{- \frac{1}{4} + \frac{1}{2} s_{i}^{\pm}}
\frac
{( a_{i}^{\frac{1}{2}} w_{i}^{- 1} q^{\frac{1}{2}  + s_{i}^{\pm}}, a_{i}^{\frac{1}{2}} w_{i} q^{\frac{1}{2}  + s_{i}^{\pm}}; q^{2} )_{\infty}}
{( a_{i}^{- \frac{1}{2}} w_{i} q^{\frac{1}{2}  + s_{i}^{\pm}}, a_{i}^{- \frac{1}{2}} w_{i}^{- 1} q^{\frac{1}{2}  + s_{i}^{\pm}}; q^{2} )_{\infty}},
\end{align}
where $s_{i}^{\pm}$ corresponds to a holonomy on $\mathbb{RP}^{2}$ and $w_{i}$ is a fugacity associated to each U$(1)_{V}$. A fugacity $a_{i}$ contains a parameter associated to each $\mathcal{P}$-type U$(1)_{A}$ symmetry as defined in single-flavor $\mathcal{N} = 4$ mirror symmetry. We rewrite this index \eqref{freeNfP} as follows:

\begin{itemize}
\item $\mathcal{P}$-SQED$_{N_{f}}$: \\ % P-SQED_{N_{f}}
To extract the diagonal part of U$(1)_{V}^{N_{f}}$, we here redefine the parameters of $\mathcal{I}_{N_{f}}^{( \mathcal{P} )}$ as
\begin{align}
	\begin{aligned}
	s_{i}^{\pm} &= m_{i}^{\pm} + m^{\pm}
	&& \text{with}
	&& \sum_{i} m_{i}^{\pm} \equiv 0 \text{ (mod $2$)}, \\ %1
	w_{i} &= z u_{i}
	&& \text{with}
	&& \prod_{i} u_{i} = 1. %2
	\end{aligned}
\end{align}
The parameters $( m^{\pm}, z )$ are ones for the diagonal part of U$(1)_{V}^{N_{f}}$ which we will denote by U$(1)_{\text{diag}}$. Gauging U$(1)_{\text{diag}}$ in this free theory make it $\mathcal{P}$-SQED$_{N_{f}}$ whose index is
\begin{align} \label{indexSQEDNfP} %the index of P-SQED_{N_{f}}
&
\mathcal{I}_{\mathcal{P}\text{-SQED}_{N_{f}}} ( q; \{ a_{i} \}; \{ m_{i}^{\pm} \}, \{ u_{i} \} ; m_{J}, \theta_{J \pm} )% \notag \\ %1
=
q^{\frac{1}{8}}
\frac{( q^{2}; q^{2} )_{\infty}}{( q; q^{2} )_{\infty}}
\sum_{m^{\pm}}
\oint
\frac{{\rm d} z}{2 \pi i z}
e^{i m^{\pm} \theta_{J \pm}}
z^{m_{J}} \notag \\ %2
&\hspace{1.5em} \times
\prod_{i = 1}^{N_{f}}
a_{i}^{- \frac{1}{4} + \frac{1}{2} ( m_{i}^{\pm} + m^{\pm} )}
\frac
{( a_{i}^{\frac{1}{2}} ( z u_{i} )^{- 1} q^{\frac{1}{2}  + m_{i}^{\pm} + m^{\pm}}, a_{i}^{\frac{1}{2}} ( z u_{i} ) q^{\frac{1}{2}  + m_{i}^{\pm} + m^{\pm}}; q^{2} )_{\infty}}
{( a_{i}^{- \frac{1}{2}} ( z u_{i} ) q^{\frac{1}{2}  + m_{i}^{\pm} + m^{\pm}}, a_{i}^{- \frac{1}{2}} ( z u_{i} )^{- 1} q^{\frac{1}{2}  + m_{i}^{\pm} + m^{\pm}}; q^{2} )_{\infty}}, %3
\end{align}
where we introduce parameters $m_{J} \in \mathbb{Z}$ and $\theta_{J \pm}$ for a $\mathcal{CP}$-type U$(1)_{J}$ topological symmetry from gauging U$(1)_{\text{diag}}$ because a background field for U$(1)_{J}$ should be coupled to a $\mathcal{P}$-type dynamical one via the BF term. This is the index for $\mathcal{P}$-SQED$_{N_{f}}$.

\item $\mathcal{CP}$-{\rm U}$(1)^{N_{f} - 1}$ gauge theory: \\ % CP-U(1)^{N_{f} - 1} gauge theory
We refer to the theory dual to $\mathcal{P}$-SQED$_{N_{f}}$ as $\mathcal{CP}$-{\rm U}$(1)^{N_{f} - 1}$ which is comprised of $( N_{f} - 1 )$ $\mathcal{CP}$-type U$(1)$ gauge groups with global symmetries corresponding to these in $\mathcal{P}$-SQED$_{N_{f}}$. Returning back to $\mathcal{I}_{N_{f}}^{( \mathcal{P} )}$ \eqref{freeNfP}, we apply the identity between \eqref{mirrorN4P-v2} and \eqref{mirrorN4P+v2} representing $\mathcal{N} = 4$ mirror symmetry to re-express  $\mathcal{I}_{N_{f}}^{( \mathcal{P} )}$ as an equivalent but quite different form $\mathcal{I}_{\mathcal{CP}\text{-U}(1)^{N_{f}}}$,
\begin{align} % the index of the dual theory_CP-U(1)^{N_{f} - 1} gauge theory
&
\mathcal{I}_{\mathcal{CP}\text{-U}(1)^{N_{f}}} ( q; \{ a_{i} \}; \{ s_{i}^{\pm} \}, \{ w_{i} \} ) \notag \\ %1
&=
\prod_{i = 1}^{N_{f}}
a_{i}^{- \frac{1}{4}}
\frac{( q, a_{i}; q^{2} )_{\infty}}{( q^{2}, a_{i}^{- 1} q; q^{2} )_{\infty}}
\sum_{s_{i} \in \mathbb{Z}}
\frac{1}{2}
\sum_{\theta_{i \pm}}
e^{i s_{i}^{\pm} \theta_{i \pm}}
w_{i}^{s_{i}}
\left( q^{\frac{1}{2}} a_{i}^{- \frac{1}{2}} \right)^{| s_{i} |}
\frac
{( e^{i \theta_{i \pm}} a_{i}^{- \frac{1}{2}} q^{| s_{i} | + 1}; q )_{\infty}}
{( e^{i \theta_{i \pm}} a_{i}^{\frac{1}{2}} q^{| s_{i} |}; q )_{\infty}}. %2
\end{align}
This is simply the index for $N_{f}$ copies of $\mathcal{CP}$-SQED$^{( \pm )}$'s having $N_{f}$ adjoint matters and $N_{f}$ $\mathcal{P}$-type U$(1)_{J}$ global symmetries whose labels are $( s_{i}^{\pm}, w_{i} )$. Here, we set new variables $( m_{i}^{\pm}, u_{i} )$ for these global symmetries such as
\begin{align} % new variables
s_{i}^{\pm} = m_{i}^{\pm} + m^{\pm},
\hspace{2em}
w_{i} = z u_{i},
\end{align}
where $( m^{\pm}, z )$ correspond to a holonomy and a Wilson line phase for the sum of $N_{f}$ $\mathcal{P}$-type U$(1)_{J}$'s which will be denoted by $\widetilde{\text{U}(1)}_{\text{diag}}$. Then, by gauging this $\mathcal{P}$-type $\widetilde{\text{U}(1)}_{\text{diag}}$, we acquire the index for $\mathcal{CP}$-{\rm U}$(1)^{N_{f} - 1}$ as
\begin{align} \label{indexdualCPv1} % the index of the dual theory_CP v1
&
\mathcal{I}_{\mathcal{CP}\text{-U}(1)^{N_{f} - 1}} ( q; \{ a_{i} \}; \{ m_{i}^{\pm} \}, \{ u_{i} \}; m_{J}, \theta_{J \pm} ) \notag \\ %1
&=
q^{\frac{1}{8}}
\frac{( q^{2}; q^{2} )_{\infty}}{( q; q^{2} )_{\infty}}
\sum_{m^{\pm}}
\oint \frac{{\rm d} z}{2 \pi i z}
e^{i m^{\pm} \theta_{J \pm}}
z^{m_{J}} \notag \\ %2
&\hspace{1em} \times
\prod_{i = 1}^{N_{f}}
a_{i}^{- \frac{1}{4}}
\frac{( q, a_{i}; q^{2} )_{\infty}}{( q^{2}, a_{i}^{- 1} q; q^{2} )_{\infty}}
\sum_{s_{i} \in \mathbb{Z}}
\frac{1}{2}
\sum_{\theta_{i \pm}}
e^{i ( m_{i}^{\pm} + m^{\pm} ) \theta_{i \pm}}
( z u_{i} )^{s_{i}}
\left( q^{\frac{1}{2}} a_{i}^{- \frac{1}{2}} \right)^{| s_{i} |}
\frac
{( e^{i \theta_{i \pm}} a_{i}^{- \frac{1}{2}} q^{| s_{i} | + 1}; q )_{\infty}}
{( e^{i \theta_{i \pm}} a_{i}^{\frac{1}{2}} q^{| s_{i} |}; q )_{\infty}}, %3
\end{align}
where $m_{J} \in \mathbb{Z}$ and $\theta_{J \pm}$ are a monopole flux and a Wilson line phase, respectively, for a topological $\widetilde{\text{U}(1)}_{J}$ which emerges from gauging $\widetilde{\text{U}(1)}_{\text{diag}}$. Now, we can evaluate the summations over $m^{\pm}$ and $z$. The sum over $m^{\pm}$ gives the delta function \eqref{mod2deltafn}
defined by mod $2$
\begin{align} \label{constraint1} % constraints from U$_{J_{g}}^{\text{sum}}$
\sum_{m^{\pm}}
e^{i m^{\pm} ( \theta_{J \pm} + \sum_{i} \theta_{i \pm} )}
=
2
\delta^{\text{mod } 2} ( \theta_{J \pm} + {\textstyle \sum_{i}} \theta_{i \pm} ),
\end{align}
and the integral over $z$ reduces to the ordinary Kronecker delta
\begin{align} \label{constraint2}
\oint \frac{{\rm d} z}{2 \pi i z}
z^{m_{J} + \sum_{i} s_{i}}
=
\delta_{m_{J} + \sum_{i} s_{i}, 0}.
\end{align}
WIth these constraints, we can split $( \theta_{i \pm}, s_{i} )$ into other (hatted) variables as follows:
\begin{align} % new variables from constraints
	\begin{aligned}
	\eqref{constraint1}
	&\Rightarrow
	\theta_{i \pm}
	=
	\hat{\theta}_{i \pm} - \hat{\theta}_{i + 1, \pm} - \frac{1}{N_{f}} \theta_{J \pm} \text{ (mod $2 \pi$)}, \\ %1
	\eqref{constraint2}
	&\Rightarrow
	s_{i}
	=
	\hat{s}_{i} - \hat{s}_{i + 1} - \frac{1}{N_{f}} m_{J}. %2
	\end{aligned}
\end{align}
Note that $m_{J}$ should be a multiple of $N_{f}$ due to gauge invariance so that $\hat{s}_{i}$ are still integers. Then, the index for the dual theory $\mathcal{CP}$-{\rm U}$(1)^{N_{f} - 1}$ can be rewritten as
\begin{align} \label{indexdualCPv2} % the index of the dual theory_CP v2
&
\mathcal{I}_{\mathcal{CP}\text{-U}(1)^{N_{f} - 1}} ( q; \{ a_{i} \}; \{ m_{i}^{\pm} \}, \{ u_{i} \}; m_{J}, \theta_{J \pm} ) \notag \\ %1
&=
q^{\frac{1}{8}}
\frac{( q^{2}; q^{2} )_{\infty}}{( q; q^{2} )_{\infty}}
\prod_{i = 1}^{N_{f}}
a_{i}^{- \frac{1}{4}}
\frac{( q, a_{i}; q^{2} )_{\infty}}{( q^{2}, a_{i}^{- 1} q; q^{2} )_{\infty}}
\sum_{\hat{s}_{i} \in \mathbb{Z}}
\frac{1}{2}
\sum_{\hat{\theta}_{i \pm}}
e^{i m_{i}^{\pm} ( \hat{\theta}_{i \pm} - \hat{\theta}_{i + 1, \pm} - \frac{1}{N_{f}} \theta_{J \pm} )}
u_{i}^{\hat{s}_{i} - \hat{s}_{i + 1} - \frac{1}{N_{f}} m_{J}} \notag \\ %2
&\hspace{1em} \times
\left( q^{\frac{1}{2}} a_{i}^{- \frac{1}{2}} \right)^{| \hat{s}_{i} - \hat{s}_{i + 1} - \frac{1}{N_{f}} m_{J} |}
\frac
{( e^{i \hat{\theta}_{i \pm} - i \hat{\theta}_{i + 1, \pm} - \frac{i}{N_{f}} \theta_{J \pm}} a_{i}^{- \frac{1}{2}} q^{| \hat{s}_{i} - \hat{s}_{i + 1} - \frac{1}{N_{f}} m_{J} | + 1}; q )_{\infty}}
{( e^{i \hat{\theta}_{i \pm} - i \hat{\theta}_{i + 1, \pm} - \frac{i}{N_{f}} \theta_{J \pm}} a_{i}^{\frac{1}{2}} q^{| \hat{s}_{i} - \hat{s}_{i + 1} - \frac{1}{N_{f}} m_{J} |}; q )_{\infty}}. %3
\end{align}
This is the precise expression\footnote{There is an additional shift symmetry acting as $\hat{s}_{i} \to \hat{s}_{i} + (\text{const.})$ for all $i$ at the same time, and we can fix one of $\hat{s}_{i}$'s by this symmetry.} of the index for $\mathcal{CP}$-{\rm U}$(1)^{N_{f} - 1}$ with showing the contribution of each bifundamental matter $\Phi_{i}$ ($i = 1, 2, \cdots, N_{f} - 1$) whose charge is $( 1, - 1 )$ only under $( \text{U}(1)_{i}, \text{U}(1)_{i + 1} )$ gauge groups in the quiver diagram.
\end{itemize}

\begin{table}[t] % Summary for symmetries
\caption{The symmetries and their fugacities in $\mathcal{P}$-SQED$_{N_{f}}$ and $\mathcal{CP}$-{\rm U}$(1)^{N_{f} - 1}$.\label{symmetries1}}
\begin{center}
\vspace{-1em}
	\begingroup
	\renewcommand{\arraystretch}{1.5}
	\begin{tabular}{|c|ll|ll|} \hline
	& \multicolumn{2}{|c}{$\mathcal{P}$-SQED$_{N_{f}}$} & \multicolumn{2}{|c|}{$\mathcal{CP}$-{\rm U}$(1)^{N_{f} - 1}$} \\ \hline %1
	\multicolumn{5}{c}{} \\[-1.9em] \hline %
	& & & & \\[-2.2em] %
	Gauge & $\mathcal{P}$-type U$(1)_{\text{diag}}$ &\hspace{-1em}: $( m^{\pm}, z )$ & $\mathcal{CP}$-type U$(1)^{N_{f} - 1}$ &\hspace{-1em}: $( \hat{s}_{i}, \hat{\theta}_{i \pm} )$ \\ \hline %2
	& & & & \\[-2em] %
	\multirow{3}{*}{Global} & $\mathcal{CP}$-type U$(1)_{J}$ &\hspace{-1em}: $( m_{J}, \theta_{J \pm} )$ & $\mathcal{CP}$-type $\widetilde{\text{U}(1)}_{J}$ &\hspace{-1em}: $( m_{J}, \theta_{J \pm} )$ \\ %4
	& $\mathcal{P}$-type U$(1)_{V}^{N_{f}} / \text{U}(1)_{\text{diag}}$ &\hspace{-1em}: $( m_{i}^{\pm}, u_{i} )$ & $\mathcal{P}$-type U$(1)_{J}^{N_{f}} / \widetilde{\text{U}(1)}_{\text{diag}}$ &\hspace{-1em}: $( m_{i}^{\pm}, u_{i} )$ \\ %5
	& $\mathcal{P}$-type U$(1)_{A}^{N_{f}}$ &\hspace{-1em}: $a_{i}$ & $\mathcal{P}$-type U$(1)_{A}^{N_{f}}$ &\hspace{-1em}: $a_{i}$ \\ \hline %6
	\end{tabular}
	\endgroup
\end{center}
\end{table}

The equality of the indices \eqref{indexSQEDNfP} for $\mathcal{P}$-SQED$_{N_{f}}$ and \eqref{indexdualCPv2} for $\mathcal{CP}$-{\rm U}$(1)^{N_{f} - 1}$ gauge theory is just Abelian $\mathcal{N} = 2$ mirror symmetry with general $N_{f}$ flavors. We summarize the symmetries and its fugacities contained in both theories in Table \ref{symmetries1}. Although all of the global symmetries are mapped in the one-to-one manner between two theories, we find that types of the parity conditions for the gauge groups \textit{are exchanged} with each other. Surely, the gauge symmetries in several theories which share the common IR fixed point do not necessarily match at UV, but this observation is an interesting, specific feature reflecting the fact that the theories are put on an unorientable manifold. Note that we can appropriately set the holonomy for U$(1)_{A}^{N_f}$ by applying the modified formula \eqref{d1loop} as mentioned above, even then the equality of the indices realizing mirror symmetry is valid.
%\begin{table}[t] % Summary for symmetries
%\caption{The symmetries and corresponding fugacities in $\mathcal{P}$-SQED$_{N_{f}}$ and $\mathcal{CP}$-{\rm U}$(1)^{N_{f} - 1}$.\label{symmetries1}}
%\vspace{-1em}
%\begin{center}
%	\begingroup
%	%\renewcommand{\arraystretch}{1.2}
%	\begin{tabular}{c|ccc} \cline{1-2} \cline{4-4}
%	& $\mathcal{P}$-SQED$_{N_{f}}$ & & $\mathcal{CP}$-{\rm U}$(1)^{N_{f} - 1}$ \\ \cline{1-2} \cline{4-4} %1
%	\multicolumn{4}{c}{} \\[-1.25em] \cline{1-2} \cline{4-4} %
%	%symmetry & type and label & & symmetry & type and label \\ \cline{1-2} \cline{4-5} %2
%	\multirow{2}{*}{Gauge} & $\mathcal{P}$-type U$(1)_{\text{diag}}$ : $( m^{\pm}, z )$ & & $\mathcal{CP}$-type U$(1)^{N_{f} - 1}$ : $( \hat{s}_{i}, \hat{\theta}_{i \pm} )$ \\[.5em] %3
%	& & & $\mathcal{P}$-type $\widetilde{\text{U}(1)}_{\text{diag}}$ : $( m^{\pm}, z )$ \\[.5em] \cline{1-2} \cline{4-4} %4
%	\multirow{3}{*}{Global} & $\mathcal{CP}$-type U$(1)_{J}$ : $( m_{J}, \theta_{J \pm} )$ & $\leftrightarrow$ & $\mathcal{CP}$-type $\widetilde{\text{U}(1)}_{J}$ : $( m_{J}, \theta_{J \pm} )$ \\[.5em] %5
%	& $\mathcal{P}$-type U$(1)_{V}^{N_{f}} / \text{U}(1)_{\text{diag}}$ : $( m_{i}^{\pm}, u_{i} )$ & $\leftrightarrow$ & $\mathcal{P}$-type U$(1)_{J}^{N_{f}} / \widetilde{\text{U}(1)}_{\text{diag}}$ : $( m_{i}^{\pm}, u_{i} )$ \\[.5em] %6
%	& $\mathcal{P}$-type U$(1)_{A}^{N_{f}}$ : $a_{i}$ & $\leftrightarrow$ & $\mathcal{P}$-type U$(1)_{A}^{N_{f}}$ : $a_{i}$ \\ \cline{1-2} \cline{4-4} %7
%	\end{tabular}
%	\endgroup
%\end{center}
%%\vspace{-1em}
%\end{table}

\paragraph{$\mathcal{CP}$-SQED$_{N_{f}}$ vs $\mathcal{P}$-{\rm U}$(1)^{N_{f} - 1}$ gauge theory.} %%%%%%%%
Let us turn to multi-flavor mirror symmetry between the SQED$_{N_{f}}$ of the $\mathcal{CP}$-type gauge symmetry and the dual quiver theory. The index $\mathcal{I}_{N_{f}}^{( \mathcal{CP} )}$ of $N_{f}$ free matters coupled to $N_{f}$ $\mathcal{C}\mathcal{P}$-type U$(1)_{V}$'s is written by
\begin{align} \label{freeNfCP} % the index of N_f free matters
\mathcal{I}_{N_{f}}^{( \mathcal{CP} )} ( q; \{ a_{i} \}; \{ s_{i} \}, \{ \theta_{i \pm} \} )
=
\prod_{i = 1}^{N_{f}}
\left( a_{i}^{- \frac{1}{2}} q^{\frac{1}{2}} \right)^{| s_{i} |}
\frac
{( e^{i \theta_{i \pm}} a_{i}^{- \frac{1}{2}} q^{1 + | s_{i} |} ; q )_{\infty}}
{( e^{i \theta_{i \pm}} a_{i}^{\frac{1}{2}} q^{| s_{i} |} ; q )_{\infty}},
\end{align}
where $s_{i}$ is a magnetic flux and $\theta_{i \pm}$ is a Wilson line phase associated to each $\mathcal{CP}$-type U$(1)_{V}$.
%where $s_{i} \in \mathbb{Z}$ is a monopole flux and $\theta_{i \pm}$ is a Wilson line phase associated to each $\mathcal{CP}$-type U$(1)_{V}$.

\begin{itemize}
\item $\mathcal{CP}$-SQED$_{N_{f}}$: \\ % CP-SQED_{N_{f}}
As done in the previous example, in order to describe $\mathcal{CP}$-SQED$_{N_{f}}$ starting from \eqref{freeNfCP}, we gauge the sum of U$(1)_{V}^{N_{f}}$ which we will name again U$(1)_{\text{diag}}$ by introducing new variables
\begin{align} % new variables
	\begin{aligned}
	s_{i} &= m_{i} + m
	&& \text{with}
	&& \sum_{i} m_{i} = 0, \\ %1
	\theta_{i \pm} &= \sigma_{i \pm} + \sigma_{\pm}
	&& \text{with}
	&& \sum_{i} \sigma_{i \pm} \equiv 0 \text{ (mod $2 \pi$)}, %2
	\end{aligned}
\end{align}
where $( m, \sigma_{\pm})$ are a monopole flux and a discretized Wilson line for $\mathcal{CP}$-type U$(1)_{\text{diag}}$. Thus, we obtain the index $\mathcal{I}_{\mathcal{CP}\text{-SQED}_{N_{f}}}$ for $\mathcal{CP}$-SQED$_{N_{f}}$ with summing over both $m$ and $\sigma_{\pm}$
\begin{align} \label{indexSQEDNfCP} % the index of \mathcal{CP}-SQED$_{N_{f}}$
&
\mathcal{I}_{\mathcal{CP}\text{-SQED}_{N_{f}}} ( q; \{ a_{i} \}; \{ m_{i} \}, \{ \sigma_{i} \}; m_{J}^{\pm}, w ) \notag \\ %1
&=
q^{- \frac{1}{8}}
\frac{( q; q^{2} )_{\infty}}{( q^{2}; q^{2} )_{\infty}}
\sum_{m \in \mathbb{Z}}
\frac{1}{2}
\sum_{\sigma_{\pm}}
e^{i m_{J}^{\pm} \sigma_{\pm}}
w^{m}
\prod_{i = 1}^{N_{f}}
\left( a_{i}^{- \frac{1}{2}} q^{\frac{1}{2}} \right)^{| m_{i} + m |}
\frac
{( e^{i ( \sigma_{i \pm} + \sigma_{\pm} )} a_{i}^{- \frac{1}{2}} q^{1 + | m_{i} + m |} ; q )_{\infty}}
{( e^{i ( \sigma_{i \pm} + \sigma_{\pm} )} a_{i}^{\frac{1}{2}} q^{| m_{i} + m |} ; q )_{\infty}}, %2
\end{align}
where $m_{J}^{\pm}$ and $w$ are a holonomy and a fugacity, respectively, associated to a topological U$(1)_{J}$ coupling to the gauged U$(1)_{\text{diag}}$ via the BF term.

\item $\mathcal{P}$-{\rm U}$(1)^{N_{f} - 1}$ gauge theory: \\ % P-U(1)^{N_{f} - 1} gauge theory
Let us call $\mathcal{P}$-{\rm U}$(1)^{N_{f} - 1}$ the quiver theory of $( N_{f} - 1 )$ $\mathcal{P}$-type U$(1)$ gauge groups with appropriate global symmetries. We can apply equivalence between \eqref{mirrorN4CP-v2} and \eqref{mirrorN4CP+v2} to rewriting the free matter index \eqref{freeNfCP}, then the resultant index $\mathcal{I}_{\mathcal{P}\text{-U}(1)^{N_{f}}}$ is
\begin{align} % the index of the dual theory_P-U(1)^{N_{f} - 1} gauge theory
&
\mathcal{I}_{\mathcal{P}\text{-U}(1)^{N_{f}}} ( q; \{ a_{i} \}; \{ s_{i} \}, \{ \theta_{i \pm} \} ) \notag \\ %1
&=
\prod_{i = 1}^{N_{f}}
a_{i}^{\frac{1}{4}}
\frac{( q^{2}, a_{i}^{- 1} q ; q^{2} )_{\infty}}{( q, a_{i} ; q^{2} )_{\infty}}
\sum_{s_{i}^{\pm}}
\oint \frac{{\rm d} z_{i}}{2 \pi i z_{i}}
e^{s_{i}^{\pm} \theta_{i \pm}}
z_{i}^{s_{i}}
\left\{
a_{i}^{- \frac{1}{4} + \frac{1}{2} s_{i}^{\pm}}
\frac
{( z_{i}^{- 1} a_{i}^{\frac{1}{2}} q^{\frac{1}{2} + s_{i}^{\pm}}, z_{i} a_{i}^{\frac{1}{2}} q^{\frac{1}{2} + s_{i}^{\pm}}; q^{2} )_{\infty}}
{( z_{i} a_{i}^{- \frac{1}{2}} q^{\frac{1}{2} + s_{i}^{\pm}}, z_{i}^{- 1} a_{i}^{- \frac{1}{2}} q^{\frac{1}{2} + s_{i}^{\pm}} ; q^{2} )_{\infty}}
\right\}. %2
\end{align}
This shows the contributions of $N_{f}$ copies of $\mathcal{N} = 4$ $\mathcal{P}$-SQED$^{( \pm )}$. To repeat the argument mentioned in Section \ref{Prescription}, we change the variables $( s_{i}, \theta_{i \pm} )$ such as
\begin{align} % new variables
s_{i} = m_{i} + m,
\hspace{2em}
\theta_{i \pm} = \sigma_{i \pm} + \sigma_{\pm},
\end{align}
where $( m, \sigma_{\pm} )$ correspond to saddle values for the sum of U$(1)_{J}^{N_{f}}$ which we will call $\widetilde{\text{U}(1)}_{\text{diag}}$. Then, gauging $\widetilde{\text{U}(1)}_{\text{diag}}$ produces the index $\mathcal{I}_{\mathcal{P}\text{-U}(1)^{N_{f} - 1}} $ of this dual theory
\begin{align} \label{indexdualPv1} % index of the dual theory_P v1
&
\mathcal{I}_{\mathcal{P}\text{-U}(1)^{N_{f} - 1}} ( q; \{ a_{i} \}; \{ m_{i} \}, \{ \sigma_{i \pm} \}; m_{J}^{\pm}, \theta_{J} ) \notag \\ %1
&=
q^{- \frac{1}{8}}
\frac{( q; q^{2} )_{\infty}}{( q^{2}; q^{2} )_{\infty}}
\sum_{m \in \mathbb{Z}}
\frac{1}{2}
\sum_{\sigma_{\pm}}
e^{i m_{J}^{\pm} \sigma_{\pm}}
w^{m}
\prod_{i = 1}^{N_{f}}
a_{i}^{\frac{1}{4}}
\frac{( q^{2}, a_{i}^{- 1} q ; q^{2} )_{\infty}}{( q, a_{i} ; q^{2} )_{\infty}} \notag \\ %2
&\hspace{1em} \times
\sum_{s_{i}^{\pm}}
\oint \frac{d z_{i}}{2 \pi i z_{i}}
e^{i s_{i}^{\pm} ( \sigma_{i \pm} + \sigma_{\pm} )}
z_{i}^{m_{i} + m}
\left\{
a_{i}^{- \frac{1}{4} + \frac{1}{2} s_{i}^{\pm}}
\frac
{( z_{i}^{- 1} a_{i}^{\frac{1}{2}} q^{\frac{1}{2} + s_{i}^{\pm}}, z_{i} a_{i}^{\frac{1}{2}} q^{\frac{1}{2} + s_{i}^{\pm}}; q^{2} )_{\infty}}
{( z_{i} a_{i}^{- \frac{1}{2}} q^{\frac{1}{2} + s_{i}^{\pm}}, z_{i}^{- 1} a_{i}^{- \frac{1}{2}} q^{\frac{1}{2} + s_{i}^{\pm}} ; q^{2} )_{\infty}}
\right\}, %3
\end{align}
where $m_{J}^{\pm}$ and $w$ also are a holonomy and a fugacity associated to U$(1)_{J}$ originated from gauging $\widetilde{\text{U}(1)}_{\text{diag}}$. Now, we focus on the consequences of gauging the parameters $m$ and $\sigma_{\pm}$. The summation over $m$ results in an usual delta function
\begin{align} \label{const4} % constraints from U$(1)_{J_{g}}^{\text{sum}}$
\sum_{m \in \mathbb{Z}}
( w {\textstyle \prod_{i}} z_{i} )^{m}
=
\delta ( w {\textstyle \prod_{i}} z_{i} ),
\end{align}
and the summation over $\sigma_{\pm}$ gives a mod $2$ delta function \eqref{mod2Kdelta}
\begin{align} \label{const3}
\frac{1}{2}
\sum_{\sigma_{\pm}}
e^{i ( m_{J}^{\pm} + \sum_{i} s_{i}^{\pm} ) \sigma_{\pm}}
=
\delta_{m_{J}^{\pm} + \sum_{i} s_{i}^{\pm}, 0}^{\text{mod } 2}.
\end{align}
As demonstrated in the previous case, to obtain the standard description of the dual theory to $\mathcal{CP}$-SQED$_{N_{f}}$, we redefine the variables by using above constraints as
\begin{align} % new variables from constraints
    \begin{aligned}
    \eqref{const4}
    &\Rightarrow
    z_{i}
    =
    \hat{z}_{i} \hat{z}_{i + 1}^{- 1} w^{- \frac{1}{N_{f}}}, \\ %1
    \eqref{const3}
    &\Rightarrow
    s_{i}^{\pm}
    =
    \hat{s}_{i}^{\pm} - \hat{s}_{i + 1}^{\pm} - \frac{1}{N_{f}} m_{J}^{\pm}
    \text{ (mod $2$)}. 
    \end{aligned}
\end{align}
Accordingly, the index for $\mathcal{P}$-{\rm U}$(1)^{N_{f} - 1}$ is written by
\begin{align} \label{indexdualPv2} % index of the dual theory_P v2
&
\mathcal{I}_{\mathcal{P}\text{-U}(1)^{N_{f} - 1}} ( q; \{ a_{i} \}; \{ m_{i} \}, \{ \sigma_{i \pm} \}; m_{J}^{\pm}, \theta_{J} ) \notag \\ %1
&=
q^{- \frac{1}{8}}
\frac{( q; q^{2} )_{\infty}}{( q^{2}; q^{2} )_{\infty}}
\prod_{i = 1}^{N_{f}}
a_{i}^{\frac{1}{4}}
\frac{( q^{2}, a_{i}^{- 1} q ; q^{2} )_{\infty}}{( q, a_{i} ; q^{2} )_{\infty}} \notag \\ %2
&\hspace{1em} \times
\sum_{\hat{s}_{i}^{\pm}}
\oint \frac{{\rm d} \hat{z}_{i}}{2 \pi i \hat{z}_{i}}
e^{i ( \hat{s}_{i}^{\pm} - \hat{s}_{i + 1}^{\pm} - \frac{1}{N_{f}} m_{J}^{\pm} ) \sigma_{i \pm}}
\left( \hat{z}_{i} \hat{z}_{i + 1}^{- 1} w^{- \frac{1}{N_{f}}} \right)^{m_{i}}
\Biggl\{
a_{i}^{- \frac{1}{4} + \frac{1}{2} ( \hat{s}_{i}^{\pm} - \hat{s}_{i + 1}^{\pm} - \frac{1}{N_{f}} m_{J}^{\pm} )} \notag \\ %3
&\hspace{1em} \times
\left.
\frac
{( \hat{z}_{i}^{- 1} \hat{z}_{i + 1} w^{\frac{1}{N_{f}}} a_{i}^{\frac{1}{2}} q^{\frac{1}{2} + \hat{s}_{i}^{\pm} - \hat{s}_{i + 1}^{\pm} - \frac{1}{N_{f}} m_{J}^{\pm}},
\hat{z}_{i} \hat{z}_{i + 1}^{- 1} w^{- \frac{1}{N_{f}}} a_{i}^{\frac{1}{2}} q^{\frac{1}{2} + \hat{s}_{i}^{\pm} - \hat{s}_{i + 1}^{\pm} - \frac{1}{N_{f}} m_{J}^{\pm}}; q^{2} )_{\infty}}
{( \hat{z}_{i} \hat{z}_{i + 1}^{- 1} w^{- \frac{1}{N_{f}}} a_{i}^{- \frac{1}{2}} q^{\frac{1}{2} + \hat{s}_{i}^{\pm} - \hat{s}_{i + 1}^{\pm} - \frac{1}{N_{f}} m_{J}^{\pm}},
\hat{z}_{i}^{- 1} \hat{z}_{i + 1} w^{\frac{1}{N_{f}}} a_{i}^{- \frac{1}{2}} q^{\frac{1}{2} + \hat{s}_{i}^{\pm} - \hat{s}_{i + 1}^{\pm} - \frac{1}{N_{f}} m_{J}^{\pm}} ; q^{2} )_{\infty}}
\right\}. %4
\end{align}
In this expression, one can see the correct contributions of bifundamental and adjoint fields attached to suitable nodes in the quiver diagram.
\end{itemize}
\begin{table}[t] % Summary for symmetries
\caption{The symmetries and thier fugacities in $\mathcal{CP}$-SQED$_{N_{f}}$ and $\mathcal{P}$-{\rm U}$(1)^{N_{f} - 1}$.\label{symmetries2}}
\begin{center}
\vspace{-1em}
	\begingroup
	\renewcommand{\arraystretch}{1.5}
	\begin{tabular}{|c|ll|ll|} \hline
	& \multicolumn{2}{|c}{$\mathcal{CP}$-SQED$_{N_{f}}$} & \multicolumn{2}{|c|}{$\mathcal{P}$-{\rm U}$(1)^{N_{f} - 1}$} \\ \hline %1
	\multicolumn{5}{c}{} \\[-1.9em] \hline %
	& & & & \\[-2.2em] %
	Gauge & $\mathcal{CP}$-type U$(1)_{\text{diag}}$ &\hspace{-1em}: $( m, \sigma_{\pm} )$ & $\mathcal{P}$-type U$(1)^{N_{f} - 1}$ &\hspace{-1em}: $( \hat{s}_{i}^{\pm}, \hat{z}_{i} )$ \\ \hline %2
	& & & & \\[-2em] %
	\multirow{3}{*}{Global} & $\mathcal{P}$-type U$(1)_{J}$ &\hspace{-1em}: $( m_{J}^{\pm}, w )$ & $\mathcal{P}$-type $\widetilde{\text{U}(1)}_{J}$ &\hspace{-1em}: $( m_{J}^{\pm}, w )$ \\ %4
	& $\mathcal{CP}$-type U$(1)_{V}^{N_{f}} / \text{U}(1)_{\text{diag}}$ &\hspace{-1em}: $( m_{i}, \sigma_{i \pm} )$ & $\mathcal{CP}$-type U$(1)_{J}^{N_{f}} / \widetilde{\text{U}(1)}_{\text{diag}}$ &\hspace{-1em}: $( m_{i}, \sigma_{i \pm} )$ \\ %5
	& $\mathcal{P}$-type U$(1)_{A}^{N_{f}}$ &\hspace{-1em}: $a_{i}$ & $\mathcal{P}$-type U$(1)_{A}^{N_{f}}$ &\hspace{-1em}: $a_{i}$ \\ \hline %6
	\end{tabular}
	\endgroup
\end{center}
\end{table}

As a result, the other example of $\mathcal{N} = 2$ multi-flavor mirror symmetry is described as the equality between the indices \eqref{indexSQEDNfCP} for $\mathcal{CP}$-SQED$_{N_{f}}$ and \eqref{indexdualPv2} for $\mathcal{P}$-{\rm U}$(1)^{N_{f} - 1}$. Table \ref{symmetries2} briefs the symmetries with variables in these thoeies. Again, though the global symmetries on one side correspond one-to-one to these on the other side, mirror symmetry accompanies the interchange of the parity type of the gauge symmetries in two theories. We note agian that the holonomy for U$(1)_{A}^{N_f}$ can be taken into account by using the modified formula \eqref{d1loop}. Since the arguments above to derive equivalence of the indices do not change in any way, we claim to find strong evidence for Abelian mirror symmetry on $\Rp^2 \times\Sp^1$ with general numbers of flavors.

%%%%%%%%%%%%%%%%%%%%%%%%%%%%%%%%%%%%%%%%%%%%%%%%%%%%%%%%
\include{common_supp_files/sonota/sections_end}
%!TEX encoding = JISJapanese

\include{common_supp_files/sonota/begin2}

%%%%%%%%%%%% Section 5 %%%%%%%%%%%%%%%%%%%%%%%%%%%%%%%%%%%%%%%
\section{Conclusion and outlook} \label{Conclusion}
% Brief summary
In this paper, we systematically construct Abelian mirror symmetry on $\Rp^2 \times \Sp^1$ with the insertions of loop operators by consistently defining the S-operation and general $N_{f}$ flavors. The unorientable structure of $\Rp^2$ is translated into mod $2$ delta functions which are analogies to the Fourier transformations of the usual delta functions on an orientable manifold. The characteristic aspect of our mirror symmetry implemented here is to interchange the two types of the parity conditions under it just as one can see in Table \ref{symmetries1}, \ref{symmetries2} for gauge symmetries of the multi-flavor theories and \eqref{gaugeW1}, \eqref{gaugeV1} for the map of the loop operators. Moreover, we should remark that the index formula \eqref{d1loop} is updated to be multiplied by the exponential prefactor from the previous version. This refinement leads to the introduction of the holonomy for U$(1)_A$ to the indices of the SQED and the XYZ model, and the statement of our mirror symmetry supported from the index viewpoint can get more strong than in the previous version.

Although we could construct consistent physical dualities on the unorientable manifold, yet there are several interesting questions which we would like to approach in the near future as listed below.

\paragraph{T-operation (Chern-Simons term).} %%%%%%%%%%%%%%%%%%%%%%%%%%%%%%%%%%
Let us comment on defining the T-operation of a SL$( 2, \mathbb{Z} )$ group \cite{Witten:2003ya}. The T-operation $T$ generates the action of adding the Chern-Simons (CS) term of a background gauge field $A$ at the level $k = 1$ to the original Lagrangian $\mathcal{L} ( \Phi )$ coupled with it:
\begin{align}
T :
\int \mathcal{D} \Phi
\ e^{\int \mathcal{L} ( \Phi ) + \int J ( \Phi ) A}
\to
\int \mathcal{D} \Phi
\ e^{\int \mathcal{L} ( \Phi ) + \int J ( \Phi ) A + \frac{1}{4 \pi} \int A \wedge {\rm d} A}.
\end{align}
Therefore, we need to have the CS term compatible under the parity action in order to set the T-operation on $\Rp^2 \times \Sp^1$ as well as the BF term for the S-operation. However, it seems impossible to obtain such T-operation because the CS term made of either a $\mathcal{P}$-type or a $\mathcal{CP}$-type vector multiplet becomes parity-odd. At the present moment, there is no appropriate complement to make it parity-even.

\paragraph{Parity anomaly.} %%%%%%%%%%%%%%%%%%%%%%%%%%%%%%%%%%%%%%%%%%
It might be related to the above comment on the T-operation, if we consider one matter multiplet coupled to a vector multiplet, then, the half-integer level Chern-Simons term for the vector multiplet emerges usually.
We do have such situation for the $\mc{P}$-type vector multiplet:
\begin{align}
\xymatrix{
*+[F-]{\Phi}
\ar@{.>}[r]%^{}_{\underbrace{}_{\ch}} 
&
*++[o][F.]{\text{\scriptsize$s^\pm , \theta$}} 
}
=
\left\{ \begin{array}{ll}
\Big(
q^{ \frac{\Delta - 1}{8}} e^{ \frac{i \theta}{4} } %\alpha^{ \frac{\fl}{4} }
\Big)^{+1}
	\frac{( e^{- i \theta} %\alpha^{- \fl} 
	q^{\frac{(2 - \Delta)}{2}} ; q^{2} )_{\infty}}{( e^{+ i \theta} %\alpha^{+ \fl}
	 q^{\frac{(0+\Delta)}{2}} ; q^{2} )_{\infty}} 
	& \text{\ for} \  (-1)^{ s^\pm} = +1, \\[.5em]
\Big(
q^{ \frac{\Delta - 1}{8}} e^{ \frac{i  \theta}{4} } %\alpha^{ \frac{\fl}{4} }
\Big)^{-1}
	\frac{( e^{- i  \theta} %\alpha^{- \fl} 
	q^{\frac{(4 - \Delta)}{2}} ; q^{2} )_{\infty}}{( e^{+ i  \theta} %\alpha^{+ \fl}
	 q^{\frac{(2+\Delta)}{2}} ; q^{2} )_{\infty}} & 
	\text{\ for}\ (-1)^{s^\pm} = -1, \\
\end{array} \right.
\label{parityanomaly}
\end{align}
which is obtained by taking $\ch_1=0$ and $\ch_2=1$ in \eqref{s1loop}.
The factors $e^{\pm \frac{i \theta}{4}}$ appearing in the so-called Casimir energy sector are indeed very similar to the Chern-Simons term contribution in the usual index \cite{Imamura:2011su}. We have no physical explanation for this ``anomalous factor" which is cancelled in all our examples with the gauge symmetry. For a global symmetry, however, our check is verified under the existence of this factor as a fugacity, hence it should have a certain physical meaning, but we now do not have any plausible interpretation. Recently, "parity" anomaly on an unorientable manifold is discussed in \cite{Witten:2016cio}, in which a similar anomalous structure to our Casimir energy $e^{\pm \frac{i \theta}{4}}$ is found. It may be interesting to study this factor.

\paragraph{3d-3d correspondence.} %%%%%%%%%%%%%%%%%%%%%%%%%%%%%%%%%%%%%%%
The geometrical aspects of the SL$( 2, \mathbb{Z} )$ action are naturally encoded in the 3d-3d correspondence \cite{Dimofte:2011ju, Dimofte:2011py} which is one interesting application of the index on $\Rp^2 \times \Sp^1$. The fundamental observation to establish this correspondence is a duality
\begin{align}
S T \circ \mathcal{T}_{1} \simeq \mathcal{T}_{1},
\end{align}
where $\mathcal{T}_{1}$ is the theory containing a free chiral multiplet (of charge $+ 1$) and the CS term of a background field at the level $k = - \frac{1}{2}$. This is derived from $\mathcal{N} = 2$ mirror symmetry with $N_{f} = 1$, and of course, we can directly check this statement by utilizing the superconformal indices. Clearly, to discuss the 3d-3d correspondence for the theories on $\Rp^2 \times \Sp^1$ based on this viewpoint, the CS term is needed to define in the parity-even way under the parity condition on our geometry. Also, we must find the correct action of the T-operation on the index consistent with parity on $\Rp^2$. Now, we do not have any resolution and would like to keep approaching this problem.

\paragraph{Factorization.} %%%%%%%%%%%%%%%%%%%%%%%%%%%%%%%%%%%%%%%%%%%
Another recent development in supersymmetric gauge theories on 3d curved manifolds is so-called \textit{factorization} \cite{Pasquetti:2011fj, Beem:2012mb, Hwang:2012jh, Taki:2013opa}, which is to factorize exact partition functions on such manifolds into basic building blocks called \textit{holomorphic blocks}. In fact, this property is related to some basic arguments to realize the 3d-3d correspondence. Thus, to find the factorised  form of our indices on $\Rp^2 \times \Sp^1$ might give us somewhat important points to establish the 3d-3d correspondence.

\paragraph{More general quiver gauge theories.} %%%%%%%%%%%%%%%%%%%%%%%%%%%%%%%%%
It is known that 3d Abelian mirror symmetry can be realized for more general quiver theories containing a number of matters charged under gauge symmetries. For instance, an U$(1)^k$ gauge theory with $N$ matter multiplets $\Phi_i$ ($i = 1, \cdots, N$) is mirror to an U$(1)^{N - k}$ gauge theory with $N$ matter multiplets $\hat{\Phi}_j$ ($i = 1, \cdots, N$) \cite{Tong:2000ky, Aganagic:2001uw}. In the former, the matters $\Phi_i$ have charges $Q_{i}^a$ ($a = 1, \cdots, k$) under the $a$-th U$(1)$ gauge symmetry, and the Chern-Simons couplings are expresed by $k^{ab} = \frac{1}{2} \sum_{i = 1}^{N} Q_{i}^a Q_{i}^b$. In the latter, the matters $\hat{\Phi}_j$ have charges $\hat{Q}_{j}^p$ ($p = 1, \cdots, N - k$) under the $p$-th U$(1)$ gauge symmetry, and the Chern-Simons couplings are given by $\hat{k}^{pq} = - \frac{1}{2} \sum_{j = 1}^{N} Q_{j}^p Q_{j}^q$. We are wondering whether we can construct such kinds of quiver theories showing mirror symmetry by appropriately connecting the $\mathcal{P}$- and $\mathcal{CP}$-type gauge symmetries.

\paragraph{Non-Abelian gauge theories.} %%%%%%%%%%%%%%%%%%%%%%%%%%%%%%%%%%%%
In addition, an extension which we should consider is to derive the index formulas of non-Abelian gauge symmetries. The difficulty to do this is lying on determining the Jacobian which arises from projecting the integration measure of the vector multiplet onto the Cartan subalgebra. The solution to this question requires us to find the precise form, especially, of the holonomy on $\Rp^2$. We hope that these will become clear and provide some new insights into the theories on unorientable manifolds.

\paragraph{Brane construction.} %%%%%%%%%%%%%%%%%%%%%%%%%%%%%%%%%%%%%%%%%
Besides the above viewpoints on the quantum field theories, it is a natural question if the brane construction in type IIB string theory \cite{deBoer:1997ka} can be constructed for mirror symmetry even on $\Rp^2 \times \Sp^1$. Two theories mirror to each other realized as the worldvolume theories on D3-branes are connected by S-duality. Since the key observation of mirror symmetry on $\Rp^2 \times \Sp^1$ is the interchange of $\mathcal{P}$-type on one side with $\mathcal{CP}$-type on the other side, we expect that this phenomena in the field theory could be accomplished by S-duality in string theory.

%%%%%%%%%%%%%%%%%%%%%%%%%%%%%%%%%%%%%%%%%%%%%%%%%%%%%%%%
\include{common_supp_files/sonota/sections_end}
\acknowledgments{We would like to thank Kentaro Hori, Yosuke Imamura, Taro Kimura, Takahiro Morimoto, Yu Nakayama, Shinsei Ryu, Masato Taki, Satoshi Yamaguchi, and Masahito Yamazaki. This research was supported by the RIKEN iTHES Project. The work of H.M. was supported in part by the JSPS Research Fellowship for Young Scientists.}
\appendix
%!TEX encoding = JISJapanese

\include{common_supp_files/sonota/begin2}

%%%%%%%%%%%% Section A %%%%%%%%%%%%%%%%%%%%%%%%%%%%%%%%%%%%%%%
\section{BPS loop operators on $\mathbb{RP}^2 \times \mathbb{S}^1$} \label{Loops}
Before proceeding to the discussion of loop operators, we note the geometrical data. The vielbein for the round two-sphere is taken to be
\begin{align} % vielbein
e^1 = d \compe, \hspace{1em}
e^2 = \sin \compe d \compt, \hspace{1em}
e^3 = d \compd.
\end{align}
The gamma matrices are denoted by $\gamma^a = \sigma^a$ ($a = 1, 2, 3$) where $\sigma^a$s are Pauli matrices, and the Killing spinors are given by
\begin{align} % Killing spinors
\e
=
e^{\frac{1}{2} ( \compd + i \compt )}
	\begin{pmatrix}
	\cos \frac{\compe}{2}
	\\[.25em]
	\sin \frac{\compe}{2}
	\end{pmatrix},
\quad
\oep
=
e^{\frac{-1}{2} ( \compd + i \compt )}
	\begin{pmatrix}
	\sin \frac{\compe}{2}
	\\[.25em]
	\cos \frac{\compe}{2}
	\end{pmatrix},
	\label{kspinor}
\end{align}
and we can easily derive their bilinear forms as
\begin{align}
( \oep \gamma_\mu \e ) 
=
	\begin{pmatrix}
	0,
	i \sin^2 \compe,
	-1
	\end{pmatrix},
\hspace{1em}
( \oep \gamma_\mu \oep ) 
=
e^{- ( \compd + i \compt )}
	\begin{pmatrix}
	- \cos \compe,
	i \sin \compe,
	- \sin \compe 
	\end{pmatrix},
\hspace{1em}
( \oep \e ) 
=
\cos \compe,
\label{bilinear}
\end{align}
where $v_\mu = ( v_\compe, v_\compt, v_\compd)$.

%%%%%%%%%%%% Section A.1 %%%%%%%%%%%%%%%%%%%%%%%%%%%%%%%%%%%%%%
\subsection{Wilson loop}

\paragraph{BPS condition.} %%%%%%%%%%%%%%%%%%%%%%%%%%%%%%%%%%%%%%%%%%%
Let us consider the Wilson loop along a curve $C$ charged under a U$(1)$ gauge symmetry,
%Let us consider the Wilson loop along a curve $C$ for a U$(1)$ global symmetry with the background vector multiplet,
\begin{align}
W_{e}^{( \pm )} ( C ) := e^{i e \int_{C} ( A_\mu \dot{x}^\mu \pm i |\dot{x}| \sigma ) d\tau},
\label{Wdef}
\end{align}
where $e$ is an electric charge. The BPS condition, $\delta_{\oep} W_{e}^{( \pm )} ( C ) =0$, which is necessary to apply localization technique (see for example \cite{Tanaka:2012nr}), turns to be
\begin{align} % BPS condition
( \oep \gamma_\mu \e ) \dot{x}^\mu \pm ( \oep \e ) |\dot{x}| = 0,
\quad
( \oep \gamma_\mu \oep ) \dot{x}^\mu = 0.
\label{BPSW}
\end{align}
It is found by simple calculation with the Killing spinors \eqref{kspinor} and the relations \eqref{bilinear} that the BPS Wilson loops wrap the $\mathbb{S}^1$ direction with sitting on the poles ($\vartheta = 0, \pi$) of the two-sphere. We express these configurations in terms of the curve $C$ as
\begin{align}
C|_{\compe =0}
&:
x^\mu = \begin{pmatrix} 0, 0, \pm \compd \end{pmatrix}, \\ %1
C|_{\compe =\pi}
&:
x^\mu = \begin{pmatrix} 0, 0, \mp \compd \end{pmatrix}, %2
\end{align}
and we get the following BPS Wilson loop localized on each pole:
\begin{align}
W_e^{( \pm )} ( C|_{\compe = 0} )
&=
e^{\pm i e \int_0^{2 \pi} d \compd ( A_{\compd} + i \sigma )|_{\compe = 0} }, \\ %1
W_e^{( \pm )} ( C|_{\compe = \pi} )
&=
e^{\mp i e \int_0^{2 \pi} d \compd ( A_{\compd} - i \sigma )|_{\compe = \pi} }. %2
\end{align}

To argue concretely the loop operators respecting the $\mathbb{Z}_2$-invariance on $\Rp^2$, we here recall the relevant parts of the $\mathcal{P}$- and $\mathcal{CP}$-type parity condition for the vector multiplet (see \cite{Tanaka:2015pwa} for the complete list):
\\
% P-type vector multiplet
\begin{minipage}{0.5\hsize}
	\begin{align}
		\begin{aligned}
 		& \text{\fbox{$\mc{P}${-type vector multiplet $V^{( \mc{P} )}$}}} \\
  		& A_{\compd}^{(\mc{P})} \to + A_{\compd}^{(\mc{P})}, \quad
 		% A_\compd \to + A_\compd,
		\si^{(\mc{P})} \to - \si^{(\mc{P})}. \\
		& \text{Locus: }
		A_\compd^{(\mc{P})} = \frac{\theta}{2 \pi}, \quad
		\si^{(\mc{P})} = 0.
  		\end{aligned} \label{vecp1} \\ \notag
	\end{align}
\end{minipage}
% CP-type Vector multiplet
\begin{minipage}{0.6\hsize}
	\begin{align}
		\begin{aligned}
		& \text{\fbox{$\mc{CP}${-type vector multiplet  $V^{( \mc{CP} )}$} }} \\
		& A_{\compd}^{(\mc{CP})} \to - A_{\compd}^{(\mc{CP})}, \quad
		% A_\compd \to + A_\compd,
		\si^{(\mc{CP})} \to + \si^{(\mc{CP})}.  \\
		& \text{Locus: }
		A_\compd^{(\mc{CP})} = \frac{\theta_\pm}{2 \pi}, \quad
		\si^{(\mc{CP})} = - s.
		\end{aligned} \label{vecp2} \\ \notag
	\end{align}
\end{minipage}

\paragraph{$\mc{P}$-type Wilson loops.} %%%%%%%%%%%%%%%%%%%%%%%%%%%%%%%%%%%%
Under the $\mc{P}$-type parity condition, the linear combination of $A_\compd$ and $\sigma$ is transformed as
\begin{align}
(A_\compd^{( \mc{P} )} + i\sigma^{( \mc{P} )})
\to
(A_\compd^{( \mc{P} )} - i\sigma^{( \mc{P} )}).
\end{align}
Thus, in order to make the operator insertion invariant under parity, it is plausible that the Wilson loop defined by $V^{( \mc{P} )}$ is taken to be a hybrid of $W_e^{( + )} (C|_{\compe = 0})$ and $W_e^{( - )} (C|_{\compe = \pi})$,
\begin{align}
W_e^{(\mc{P})}
&=
W_e^{( + )} (C|_{\compe = 0})
\cdot
W_e^{( - )} (C|_{\compe = \pi})
\notag \\ %1
%&=
%\exp \left[ i e \int_0^{2 \pi} d \compd ( A_{\compd}^{( \mc{P} )} + i \sigma^{( \mc{P} )} )|_{\compe =0} \right]
%\cdot
%\exp \left[ i e \int_0^{2 \pi} d \compd ( A_{\compd}^{( \mc{P} )} - i \sigma^{( \mc{P} )} )|_{\compe =\pi} \right]
%\notag \\ %2
&=
\exp \left[ 2 i e \int_0^{2 \pi} d \compd A_{\compd}^{( \mc{P} )}|_{\compe = 0, \pi} \right] \notag \\ %3
&\overset{\text{Locus}}{\longrightarrow}
e^{2 i e \theta}. %4
\label{PWil}
\end{align}
We can interpret it from the index point of view \eqref{indexP} just by ungauging it, and we denote this as somewhat ``trivial" insertion of the Wilson loop associated with a $\mc{P}$-type global symmetry as if it is an operation on the index with background:
\begin{align}
\hat{w}_{e}
\cdot
\mathcal{I} ( q; s^{\pm}, \theta )
&:=
e^{2 i e \theta}
\mathcal{I} ( q; s^{\pm}, \theta ).
%\notag \\ %1
%\Longleftrightarrow
%\Bigg(
%\xymatrix{
%\dots
%\ar@{->}[r]_>{\qquad \qquad \qquad \quad W_e}
%&
%*++[o][F.]{\text{\scriptsize$s^\pm , \theta$}} 
%}
%\Bigg)
%&=
%e^{2 i e \theta} 
%\Bigg(
%\xymatrix{
%\dots
%\ar@{->}[r]
%&
%*++[o][F.]{\text{\scriptsize$s^\pm , \theta$}} 
%}
%\Bigg). %2
\end{align}
Then, the gauge Wilson loop insertion associated with the $\mc{P}$-type dynamical vector multiplet is represented by
\begin{align}
\hat{W}_e \cdot \mc{I} (q)
=
q^{+\frac{1}{8} }
\frac{(q^2;q^2)_\infty}{(q;q^2)_\infty}
\sum_{s^\pm =0,1}
\frac{1}{2\pi} \int_0^{2\pi} d \theta  \
\hat{w}_{e}
\cdot
\mathcal{I} ( q; s^{\pm}, \theta ).
\label{Wandw}
\end{align}

\paragraph{$\mc{CP}$-type Wilson loops.} %%%%%%%%%%%%%%%%%%%%%%%%%%%%%%%%%%%%
In contrast, the linear combination of $A_\compd$ and $\sigma$ is transformed under the $\mc{CP}$-type parity condition such that
\begin{align}
( A_\compd^{( \mc{CP} )} + i \sigma^{( \mc{CP} )})
\to
- ( A_\compd^{( \mc{CP} )} - i \sigma^{( \mc{CP} )}).
\end{align}
This observation implies that the Wilson loop defined by $V^{( \mc{CP} )}$ should be comprised\footnote{For later use, we choose a \textit{pure imaginary half-integer} charge.} of $W_{i e/2}^{( + )} ( C|_{\compe = 0} )$ and $W_{- i e/2}^{( - )} ( C|_{\compe = \pi} )$ to cancel the nontrivial change of the sign,
\begin{align}
W_{e}^{(\mc{CP})}
&=
W_{i e/2}^{( + )} ( C|_{\compe = 0} )
\cdot
W_{- i e/2}^{( - )} ( C|_{\compe = \pi} )
\notag \\ %1
%&=
%\exp \left[ - \frac{e}{2} \int_0^{2 \pi} d \compd ( A_{\compd}^{( \mc{CP} )} + i \sigma^{( \mc{CP} )} )|_{\compe = 0} \right]
%\cdot
%\exp \left[ + \frac{e}{2} \int_0^{2 \pi} d \compd ( A_{\compd}^{( \mc{CP} )} - i \sigma^{( \mc{CP} )} )|_{\compe =\pi} \right]
%\notag \\ %2
&=
\exp \left[ - i e \int_0^{2 \pi} d \compd \ \sigma^{( \mc{CP} )} |_{\compe = 0, \pi} \right] \notag \\ %3
&\overset{\text{Locus}}{\longrightarrow}
e^{ 2 \pi i e s}. %4
\label{CPWil}
\end{align}
We can interpret it from the index point of view \eqref{indexCP} just by ungauging it, and we again denote this as somewhat ``trivial" insertion of the Wilson loop associated with a $\mc{CP}$-type global symmetry as if it is an operation on the index with background:
\begin{align}
\hat{w}_{e}
\cdot
\mathcal{I} ( q; s, \theta_{\pm} )
&=
e^{2 \pi i e s}
\mathcal{I} ( q; s, \theta_{\pm} ).
%\notag \\ %1
%\Longleftrightarrow
%\Bigg(
%\xymatrix{
%\dots
%\ar@{->}[r]<1mm>
%\ar@{<-}[r]<-1mm>_>{\qquad \qquad \qquad \ W_e}
%&
%*++[o][F.]{\text{\scriptsize$s , \theta_\pm$}} 
%}
%\Bigg)
%&=
%e^{2 \pi i e s} 
%\Bigg(
%\xymatrix{
%\dots
%\ar@{->}[r]<1mm>
%\ar@{<-}[r]<-1mm>
%&
%*++[o][F.]{\text{\scriptsize$s , \theta_\pm$}} 
%}
%\Bigg). %2
\end{align}
Then, the gauge Wilson loop insertion associated with $\mc{CP}$-type dynamical vector multiplet is represented by
\begin{align}
\hat{W}_e \cdot \mc{I} (q)
=
q^{-\frac{1}{8} }
\frac{(q;q^2)_\infty}{(q^2;q^2)_\infty}
\sum_{s \in \mathbb{Z} }
\frac{1}{2} \sum_{\theta_\pm = 0, \pi} \
\hat{w}_{e}
\cdot
\mathcal{I} ( q; s, \theta_\pm ).
\label{Wandw2}
\end{align}

\paragraph{S-transform point of view.} %%%%%%%%%%%%%%%%%%%%%%%%%%%%%%%%%%%%%%
More generically, we can turn on the background field coupled to a topological symmetry when we consider the gauge theory, then the relations \eqref{Wandw} and \eqref{Wandw2} are promoted to the following relation:
\begin{align}
\hat{W}_{e}
=
\hat{S} \hat{w}_e
.
\label{WandSandw}
\end{align}

%\newpage
%%%%%%%%%%%% Section A.2 %%%%%%%%%%%%%%%%%%%%%%%%%%%%%%%%%%%%%%
\subsection{Vortex loop}

\paragraph{Definition via S-operation.} %%%%%%%%%%%%%%%%%%%%%%%%%%%%%%%%%%%%%%
As studied in \cite{Kapustin:2012iw}, a vortex loop operator associated with a global symmetry can be defined by using the S-operation which we explain in Section \ref{cal},
\begin{align}
\hat{v} _m = \hat{S}^{-1} \hat{w}_m \hat{S},
\label{Vdef}
\end{align}
where $m$ is a magnetic charge of the gauge symmetry. This means that the BPS vortex loop also extends to $\Sp^1$ and is located at the poles of the two-sphere, which has been directly derived from the boundary condition preserving supersymmetry in \cite{Kapustin:2012iw, Drukker:2012sr}. Now, it is quite convenient to accept this definition in constructing the $\mathbb{Z}_2$-invarient vortex loops on $\Rp^2 \times \Sp^1$ because we already saw that the S-operation does work well.

%\vspace{.5cm}\noindent
For the vortex loop associated with the gauge symmetry is defined by 
\begin{align}
\hat{V}_m = \hat{S} \hat{v}_m
\end{align}
as observed in the Wilson loop \eqref{WandSandw}. By this construction, the vortex loop automatically satisfies $\mathbb{Z}_2$-invariance and the BPS condition. Thus, we focus on the vortex loop for the global symmetry here.

\paragraph{$\mc{P}$-type vortex loops.} %%%%%%%%%%%%%%%%%%%%%%%%%%%%%%%%%%%%%
The $\mc{P}$-type vortex loop operator acts on the index \eqref{indexP} coupled to a $\mc{P}$-type global symmetry,
\begin{align}
\hat{v}_m \cdot \mc{I}^{\Rp^2} ( q; s^\pm, \theta ).
%=
%\Big(
%\xymatrix{
%\dots
%\ar@{->}[r]_>{\qquad \qquad \qquad \quad v_{m}}
%&
%*++[o][F.]{\text{\scriptsize$s^\pm , \theta$}} 
%}
%\Big).
\end{align}
We can translate this in terms of the $\mc{CP}$-type Wilson loop \eqref{CPWil} by using \eqref{Vdef}, and it is evaluated with the rules of the S-operation \eqref{defS} so that the insertion of $v_m$ changes the Wilson line phase $\theta$ by $2 \pi$ times the magnetic charge,

\begin{align}
&
\hat{v}_m \cdot \mc{I}^{\Rp^2} ( q; s^\pm, \theta ) \notag \\ %1
&=
\Big(
\xymatrix{
\dots
\ar@{->}[r]
&
*++[o][F-]{\text{\scriptsize$\mc{P}$}} 
\ar@{-}[r]^{\text{BF}}
&
*++[o][F-]{\text{\scriptsize$\mc{CP}$}} 
\ar@{.}[r]^{-\text{BF}}_<{\quad \ \ W_{m}}
&
*++[o][F.]{\text{\scriptsize$s^\pm, \theta$}} 
}
\Big)
\notag \\ %2
&=
\underbrace{
%\left[
\sum_{s_1^\pm} 
\int_0^{2\pi} \frac{d \theta_1}{2 \pi}
q^{\frac{1}{8}} \frac{( q^2; q^2 )_\infty}{( q; q^2 )_\infty}
\mc{I}^{\Rp^2} ( q; s_1^\pm, \theta_1 )
%\right]
}_{
                    \xymatrix {
                    \dots
                         \ar@{->}[r]
                         &
                    *++[o][F-]{\text{\scriptsize$\mc{P}$}} 
                    }
 }
\underbrace{
%\left\{
\sum_{s_2 \in \mathbb{Z}}
\frac{1}{2} \sum_{\theta{_2}{_\pm}}
q^{- \frac{1}{8}} \frac{( q; q^2 )_\infty}{( q^2; q^2 )_\infty}
}_{
                    \xymatrix {
                    *++[o][F-]{\text{\scriptsize$\mc{CP}$}} 
                    }
}
\underbrace{
%\left(
e^{i s^\pm_1 \theta{_2}{_\pm}}
e^{i {s_2} \theta_1}
%\right)
}_{\text{BF}}
\cdot
\underbrace{
e^{2 \pi i m s_2}
}_{W_m^{(\mc{CP})}}
\cdot
\underbrace{
%\left(
e^{- i s^\pm \theta{_2}{_\pm}}
e^{- i s_2 \theta}
%\right)
%\right\}
}_{-\text{BF}}
\notag \\ %3
&=
\sum_{s^\pm_1} 
\int_0^{2 \pi} \frac{d \theta_1}{2 \pi}
\mc{I}^{\Rp^2} ( q; s_1^\pm, \theta_1 )
\left\{
\sum_{s_2 \in \mathbb{Z}}
e^{i s_2 ( \theta_1  - \theta + 2 \pi m )}
\right\}
\left\{
\frac{1}{2} \sum_{\theta{_2}{_\pm}}
e^{i ( s^\pm_1 - s^\pm ) \theta{_2}{_\pm}}
\right\}
\notag \\ %4
&=
\sum_{s^\pm_1} 
\int_0^{2 \pi} \frac{d \theta_1}{2 \pi}
\mc{I}^{\Rp^2} ( q; s_1^\pm, \theta_1 )
\biggl\{
2 \pi
\delta ( \theta_1  - \theta + 2 \pi m )
\biggr\}
\biggl\{
\delta_{ s^\pm_1 , s^\pm }^{\text{mod 2}}
\biggr\}
\notag \\ %5
&=
\mc{I}^{\Rp^2} ( q; s^\pm,  \theta - 2 \pi m ) \notag \\ %6
&=
e^{- 2 \pi m \frac{\pa}{\pa \theta}}
\mc{I}^{\Rp^2} ( q; s^\pm,  \theta ). %7
\end{align}
Note that, if $m \in \mathbb{Z}$, it is identical to the original index because $\theta$ is defined modulo $2 \pi$. To make it easy to use this insertion in our applications, we write the insertion of the $\mc{P}$-type vortex loop with a U$(1)$ global symmetry as the second quiver diagram in \eqref{VP}.

\paragraph{$\mc{CP}$-type vortex loops.} %%%%%%%%%%%%%%%%%%%%%%%%%%%%%%%%%%%%
Further, we can take the vortex loop acting on the index \eqref{indexCP},
%In addition, we can take the $\mc{CP}$-type vortex loop acting on the index \eqref{indexCP} with a certain $\mc{CP}$-type global symmetry,
\begin{align}
\hat{v}_m \cdot \mc{I}^{\Rp^2} ( q; s, \theta^\pm ).
%=
%\Big(
%\xymatrix{
%\dots
%\ar@{->}[r]<1mm>
%\ar@{<-}[r]<-1mm>_>{\qquad \qquad \qquad \ v_{m}}
%&
%*++[o][F.]{\text{\scriptsize$s , \theta_\pm$}} 
%}
%\Big).
\end{align}
Unlike the previous case, the definition \eqref{Vdef} with \eqref{PWil} and repeating \eqref{defS} leads to the exact action that the magnetic flux $s$ is shifted by a doubled magnetic charge:
\begin{align}
&
\hat{v}_m \cdot \mc{I}^{\Rp^2} ( q; s, \theta^\pm ) \notag \\ %1
&= 
\Big(
\xymatrix{
\dots
\ar@{->}[r]<1mm>
\ar@{<-}[r]<-1mm>
&
*++[o][F-]{\text{\scriptsize$\mc{CP}$}} 
\ar@{-}[r]^{\text{BF}}
&
*++[o][F-]{\text{\scriptsize$\mc{P}$}} 
\ar@{.}[r]^{-\text{BF}
}_<{\quad W_{m}}
&
*++[o][F.]{\text{\scriptsize$s, \theta^\pm$}} 
}
\Big)
\notag \\ %2
&=
\underbrace{
\sum_{s_1 \in  \mathbb{Z}}
\frac{1}{2}
\sum_{\theta_1^\pm}
q^{- \frac{1}{8}} \frac{( q; q^2 )_\infty}{( q^2; q^2 )_\infty}
\mc{I}^{\Rp^2} ( q; s_1, \theta_1^\pm )
}_
{
\xymatrix {
                    \dots
                         \ar@{->}[r]<1mm>
                         \ar@{<-}[r]<-1mm>
&
                    *++[o][F-]{\text{\scriptsize$\mc{CP}$}} 
                    }
}
\underbrace{
%\left\{
\sum_{s_2^\pm}
\int_0^{2 \pi} \frac{d \theta_2}{2 \pi}
q^{\frac{1}{8}} \frac{( q^2; q^2 )_\infty}{( q; q^2 )_\infty}
}_
{
\xymatrix {
                    *++[o][F-]{\text{\scriptsize$\mc{P}$}} 
                    }
}
\underbrace{
%\left(
e^{i s_2^\pm \theta{_1}{_\pm}}
e^{i s_1 \theta_2}
}_{\text{BF}}
%\right)
\cdot
\underbrace{
e^{2 i m \theta_2}
}_{W_m^{(\mc{P})}}
\cdot
\underbrace{
%\left(
e^{- i s_2^\pm \theta_\pm}
e^{- i s \theta_2}
%\right)
}_{-\text{BF}}
%\right\}
\notag \\ %3
&=
\sum_{s_1 \in  \mathbb{Z}}
\frac{1}{2}
\sum_{\theta_1^\pm}
\mc{I}^{\Rp^2} ( q; s_1, \theta{_1}{_\pm} )
\left\{
\sum_{s_2^\pm}
e^{i s_2^\pm ( \theta{_1}{_\pm} - \theta_\pm )}
\right\}
\left\{
\int_0^{2 \pi} \frac{d \theta_2}{2 \pi}
e^{i ( s_1 + 2  m - s ) \theta_2}
\right\}
\notag \\ %4
&=
\sum_{s_1 \in  \mathbb{Z}}
\frac{1}{2}
\sum_{\theta{_1}{_\pm}}
\mc{I}^{\Rp^2} ( q; s_1, \theta{_1}{_\pm} )
\biggl\{
2 \delta^{\text{mod 2}} ( \theta{_1}{_\pm}  -  \theta_\pm )
\biggr\}
\biggl\{
\delta_{ s_1 , s - 2 m}
\biggr\}
\notag \\ %5
&=
\mc{I}^{\Rp^2} ( q; s - 2m, \theta_\pm ) \notag \\ %6
&=
e^{- 2 m \frac{\pa}{\pa s}}
\mc{I}^{\Rp^2} ( q; s , \theta_\pm ). %7
\end{align}
In conclusion, the quiver rule for this $v_{m}$ insertion is shown as the second diagram in \eqref{VCP}.

%%%%%%%%%%%%%%%%%%%%%%%%%%%%%%%%%%%%%%%%%%%%%%%%%%%%%%%%
\include{common_supp_files/sonota/sections_end}

\providecommand{\href}[2]{#2}\begingroup\raggedright\endgroup

\end{document}